\documentclass[journal=nalefd,manuscript=paper,layout=twocolumn,manuscript=letter]{achemso}
\setkeys{acs}{articletitle=true}
\setkeys{acs}{etalmode=truncate}
\setkeys{acs}{maxauthors=10}
\usepackage{color}
 \usepackage{xcolor}

\makeatletter
\let\l@addto@macro\relax
\makeatother
\usepackage[fontsize=11 pt]{scrextend}
\usepackage[version=3]{mhchem} 
 
\usepackage{mciteplus}
\usepackage{amsmath}
\usepackage{cuted}
\usepackage{bbold}
\usepackage{mathtools, cuted}
 \usepackage[utf8]{inputenc}

\newsavebox{\varmatrixbox}

\title[An \textsf{achemso} demo]{ Microscopic Theory of Multimode Polariton Dispersion in Multilayered Materials}

\author{Arkajit Mandal}%
\email{am5813@columbia.edu}
\affiliation{\small Department of Chemistry, Columbia University, 3000 Broadway, New York, New York, 10027,  U.S.A}
\author{Ding Xu}%
\affiliation{\small Department of Chemistry, Columbia University, 3000 Broadway, New York, New York, 10027,  U.S.A}

\author{Ankit Mahajan}%
\affiliation{\small Department of Chemistry, Columbia University, 3000 Broadway, New York, New York, 10027,  U.S.A}

\author{Joonho Lee}%
\affiliation{\small Department of Chemistry, Columbia University, 3000 Broadway, New York, New York, 10027,  U.S.A}

\author{Milan E. Delor}%
\affiliation{\small Department of Chemistry, Columbia University, 3000 Broadway, New York, New York, 10027,  U.S.A}

\author{David R. Reichman}
\email{drr2103@columbia.edu}
\affiliation{\small Department of Chemistry, Columbia University, 3000 Broadway, New York, New York, 10027,  U.S.A}

\begin{document}


\begin{abstract}{\footnotesize
We develop a microscopic theory for the multimode polariton dispersion in materials coupled to cavity radiation modes. Starting from a microscopic light-matter Hamiltonian, we devise a general strategy for obtaining simple matrix models of polariton dispersion curves based on the structure and spatial location of multi-layered 2D materials inside the optical cavity. Our theory exposes the connections between seemingly distinct models that have been employed in the literature and resolves an ambiguity that has arisen concerning the experimental description of the polaritonic band structure. We demonstrate the applicability of our theoretical formalism by fabricating various geometries of multi-layered perovskite materials coupled to cavities and demonstrating that our theoretical predictions agree with the experimental results presented here. 
}
\end{abstract}

\maketitle

{\hfill \break
\hfill \break }
 \begin{figure}
\centering
\includegraphics[width=1.0\linewidth]{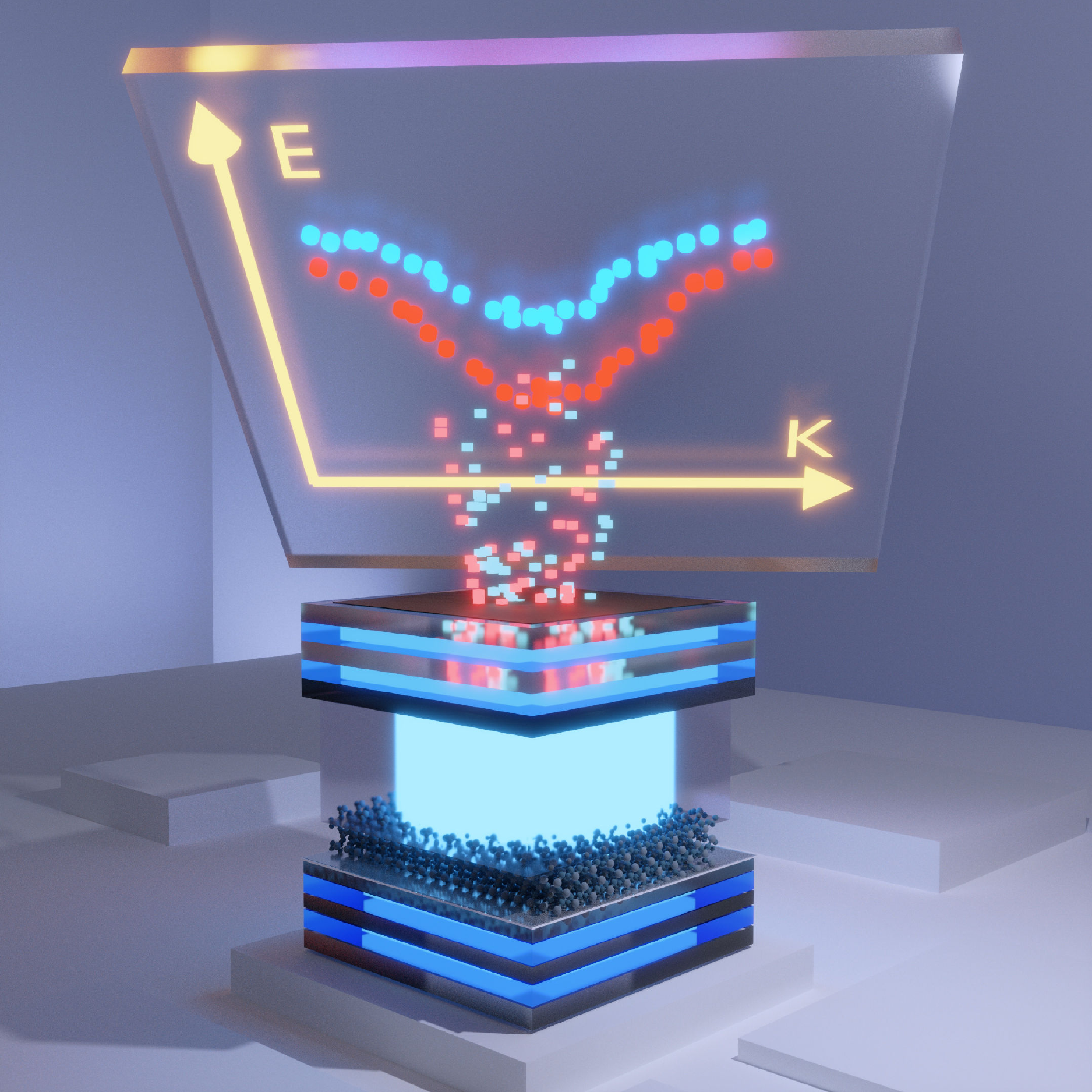}
\end{figure}

{\footnotesize
\hfill \break
 {\bf Introduction.} Strongly coupling matter to quantized radiation via an optical cavity enables the generation of exciting new chemical~\cite{Hutchison2012ACIE,Feist2018AP, Thomas2019S,GarciaVidal2021S,Nagarajan2021JACS,Ribeiro2018CS,Mandal2020JPCB,Semenov2019JCP,Weijun2022JCP,Mandal2022CR,Li2021ARPC} and physical phenomena~\cite{Berghuis2022AP,Xu2022,Deng2010RMP,Kockum2019NRP,Keeling2020ARPC,Arnardottir2020PRL,Rozenman2018ACSP,Sanchez2020JPCL,Mandal2020JPCL} in a highly controllable manner. 
Polaritons, light-matter hybrid quasi-particles, are formed under light-matter coupling strengths that are larger than competitive dissipative processes such as cavity loss~\cite{Barquilla2022ACSP,Reithmaier2004N,Muller2015PRX,Laussy2012JP}. 
The properties of polaritons are readily characterized in absorption or photoluminescence spectra, by Rabi-splitting features, and by new dispersion with an effective mass much \textit{smaller} than matter~\cite{Deng2003PNAS,Kockum2019NRP,Keeling2020ARPC}. 
Despite decades of research on polaritons, many aspects of polariton physics remain elusive~\cite{Georgiou2021JCP,Richter2015APL}.

The polariton dispersion in a Fabry–P\'{e}rot cavity for a single mode   coupled to a one-dimensional excitonic chain model with orientation parallel to the cavity mirrors, is obtained by diagonalizing a simple $2\times2$ matrix~\cite{Michetti2005PRB,Tichauer2021JCP,Gerace2007PRB,Georgiou2021JCP,Richter2015APL}. The diagonal elements of this $2\times2$ matrix (one for the photon and another for the exciton) correspond to the uncoupled excitonic and photonic energies at a particular longitudinal wavevector, and the off-diagonal terms capture the light-matter coupling. When  considering $N$ cavity modes coupled to an exciton chain, one may extend the  $2\times2$  matrix to an $(N+1)\times (N+1)$ matrix where the single exciton couples to all $N$ cavity modes. The experimentally obtained polariton dispersion, on the other hand, often deviates from the predictions of the $(N+1)\times (N+1)$ model and instead is better described by a $2N\times 2N$ model~\cite{Xu2022,Richter2015APL,Georgiou2021JCP,Balasubrahmaniyam2021PRB}. This $2N\times 2N$ model matrix is constructed by making $N$ copies of the exciton branch where each exciton branch couples to one cavity mode branch, such that the overall matrix is block diagonal with $N$ $2\times2$ subblocks. Previously, classical Maxwell theory has been used to investigate the polariton dispersion, where the $2N\times 2N$ model appears to predict the observed polariton dispersion correctly.~\cite{Balasubrahmaniyam2021PRB,Richter2015APL} However, these studies do not provide a full microscopic understanding of these effects, the origin of the $2N\times 2N$ model from the microscopic light-matter Hamiltonian, or discuss its validity in relation to the spatial geometry of the material.    

In this work, we develop a quantum mechanical microscopic theory to understand and predict the multi-mode polariton dispersion of multi-layered materials coupled to radiation in a Fabry–P\'{e}rot cavity. In the following, we develop a general $(N+N_e)\times (N+N_e)$ model (with $N_e \le N$ exciton branches) that, for specific geometries and spatial locations of multi-layered materials, reduces to $(N+2)\times(N+2)$, $2N\times 2N$, or $(N+1)\times(N+1)$ models. We then show that the $(N+1)\times (N+1)$ model, which is derived for a single-layer material, cannot be directly used for multi-layered materials often studied in experiments. We  demonstrate that regardless of inter-layer coupling, for filled cavities, the $2N\times 2N$ model is appropriate. Finally, we show the applicability of this theoretical formalism by preparing multi-layered perovskite materials coupled to cavities with various spatial geometries inside the cavity. We show that the multimode polariton dispersion predicted by our theoretical model agrees reasonably well with the   experimental results provided here. 

  }
 
{\footnotesize

{\bf Exciton-Polariton Hamiltonian.} Here we consider a generalized Tavis-Cummings~\cite{Keeling2020ARPC, Mandal2022CR, Keeling2012} (GTC)  Hamiltonian describing a (Frenkel) exciton-polariton system beyond the usual long-wavelength approximation~\cite{Keeling2012, Mandal2022CR, Keeling2020ARPC, JiajunPRB2020, DmytrukPRB2021}, which we rigorously derive from the p.A Hamiltonian using orbital and nuclear-centered gauge transformations, with details provided in the Supporting Information (SI). The exciton-polariton Hamiltonian of a multi-layered material in a cavity with cavity quantization along $y$ is given as 

\begin{align}\label{TC-k.x}\nonumber
\mathcal{\hat{H}}_\mathrm{GTC}(k_z=0)
&= \sum_{k_x} 
\Bigg( 
\sum_{k_y}\hat{a}_{\boldsymbol k}^{\dagger}\hat{a}_{\boldsymbol k}\omega_c({\boldsymbol k}) + \sum_{m}\epsilon(k_x)\hat{d}^{\dagger}_{k_x,m}\hat{d}_{k_x,m} \\ \nonumber
&+ \sum_{{k_y},m}g_{\boldsymbol k}   \big(  \hat{a}_{{\boldsymbol k}}^\dagger \hat{d}_{k_x,m} +    \hat{a}_{{\boldsymbol k}} \hat{d}_{ k_x,m}^{\dagger} \big)\sin({ k_y}  Y_{m})\Bigg) \\
&\equiv\sum_{k_x} \hat{h}_\mathrm{GTC}(k_x),
\end{align}

\noindent where $g_{\boldsymbol k} = \sqrt{N_x N_z}\mu_0 \lambda_{\boldsymbol k} $ and
 $\epsilon(k_x) = \omega_{0} - 2\tau_z- 2\tau_x\cos(k_x  \delta l_x)$ for a simple nearest neighbor coupling $\tau_x$ and $\tau_z$ along $x$ and $z$, $\hat{d}_{k_x, m}^{\dagger}$ creates a material excitation in the $m$th layer located at $Y_m$ with in-plane (with respect to the mirrors) wavevector $k_x$, $\hat{a}_{\boldsymbol k}^{\dagger}$ is the photonic creation operator of the cavity mode  ${\boldsymbol{k}} = (k_x, k_y, k_z)$ with photon frequency $\omega_{\boldsymbol{k}} = \frac{c }{\eta}|{\boldsymbol{k}}|$ ($c$ is the speed of light and $\eta$ is the refractive index) such that $k_x = \frac{2\pi}{L_x}n_x$  with  $n_x = 0, \pm 1, \pm  2, ...$ with $L_x$ as the length of the periodic super-cell along $x$ direction with a similar expression for $k_z$~\cite{Tichauer2021JCP}. In the main text we have set $k_z = 0$ for both matter and cavity, as in most experiments (including ours) the polariton dispersion is plotted along $k_x$ while $k_z$ is set to 0. Further, ${ \boldsymbol \lambda}_{{\boldsymbol k}} = \sqrt{\frac{\hbar \omega_c({\boldsymbol k})}{{\epsilon_0  \epsilon_r  \mathcal{V}}}} \hat{\boldsymbol e}_{{\boldsymbol k}}$, where $\epsilon_0$  and $ \epsilon_r $ are the vacuum and material permittivity, respectively, $\mathcal{V} = L_x L_y L_z$ is the quantization volume (equivalently the super-cell volume containing $N_x N_y N_z$ unit cells), and $\hat{\boldsymbol e}_{{\boldsymbol k}} \perp  {\boldsymbol k}$ is the polarization direction of the radiation mode ${\boldsymbol k}$.  In our model, the cavity mirror impose a boundary condition along the $y$ direction and thus $k_y = \frac{\pi}{L_y}n_x$  with  $n_y = 0, \pm 1, \pm  2, ..., \pm n_{y_\text{max}}$ ($n_{y_\text{max}}$ is a numerical cut-off ).

The GTC Hamiltonian is block-diagonal in each $k_x$ subspace containing only $\{\hat{a}_{k_x,k_y},   \hat{d}_{k_x,m}\}$
and their Hermitian conjugates with ${\hat{h}}_\mathrm{GTC}(k_x)$ as the Hamiltonian in the $k_x$ block. Despite its convenience, an undesirable feature of the GTC Hamiltonian is that it does not converge with respect to the number of cavity modes. In the SI, we show that this is due to the absence of the dipole-self energy term in the GTC Hamiltonian and that, perhaps counterintuitively, the GTC Hamiltonian produces accurate results only when considering a small number of energetically relevant cavity modes.  


{\bf General strategy for multi-layered materials.} For materials with an arbitrary thickness placed inside a cavity, we develop a general strategy for obtaining the polariton dispersion based on defining new matter operators that take advantage of the structure of the light-matter coupling. Consider a total multi-layer width of $l_y = \delta l_y \cdot N_y$ (with inter-layer distance $\delta  l_y$), where $l_y \le L_y$. Considering $N$ energetically relevant cavity mode branches such that we restrict  $n_\mathrm{min}\frac{\pi}{\eta L_y} < k_y < n_\mathrm{max}\frac{\pi}{\eta L_y}$ (below we assume the refractive index $\eta$ to be unity unless noted otherwise), we construct the following  $N_y \times N_e$ matrix (where $N_e \le N = n_\mathrm{max} - n_\mathrm{min}$) using the cavity mode functions as 
 
\begin{align}\small
\Lambda_{\mathrm{NO}}  =  
\begin{bmatrix}
\sin (\frac{\pi n_1}{L_y}   Y_1 ) & \sin (\frac{\pi n_2}{L_y}   Y_1 ) & \hdots \\
\sin (\frac{\pi n_1    }{L_y}   Y_2 ) & \sin (\frac{\pi n_2  }{L_y}     Y_2 ) &\hdots \\
\vdots & \vdots & \hdots \\
\sin (\frac{\pi n_1}{L_y}   Y_{N_{y}} ) & \sin (\frac{\pi n_2}{L_y}     Y_{N_{y}} )  & \hdots
\end{bmatrix}\label{NO-mat}
\end{align}
where $\Lambda_{\mathrm{NO}}$  is generally a  non-orthogonal matrix. Note that here $\{ n_1, n_2, ... \}$ are chosen such that the above $N_y \times N_e$ matrix contains linearly independent columns. In most cases, $N_e = N = n_\mathrm{max} - n_\mathrm{min}$, so that relevant cavity mode functions are linearly independent.  We numerically construct the $\Lambda_{\mathrm{NO}}$ matrix by first initializing it with a single column and then adding a column only if the rank (defined as the number of non-zero singular values) of the matrix containing this additional column increases by one. 
 \begin{figure*}[!t]
\centering
\includegraphics[width=0.95\linewidth]{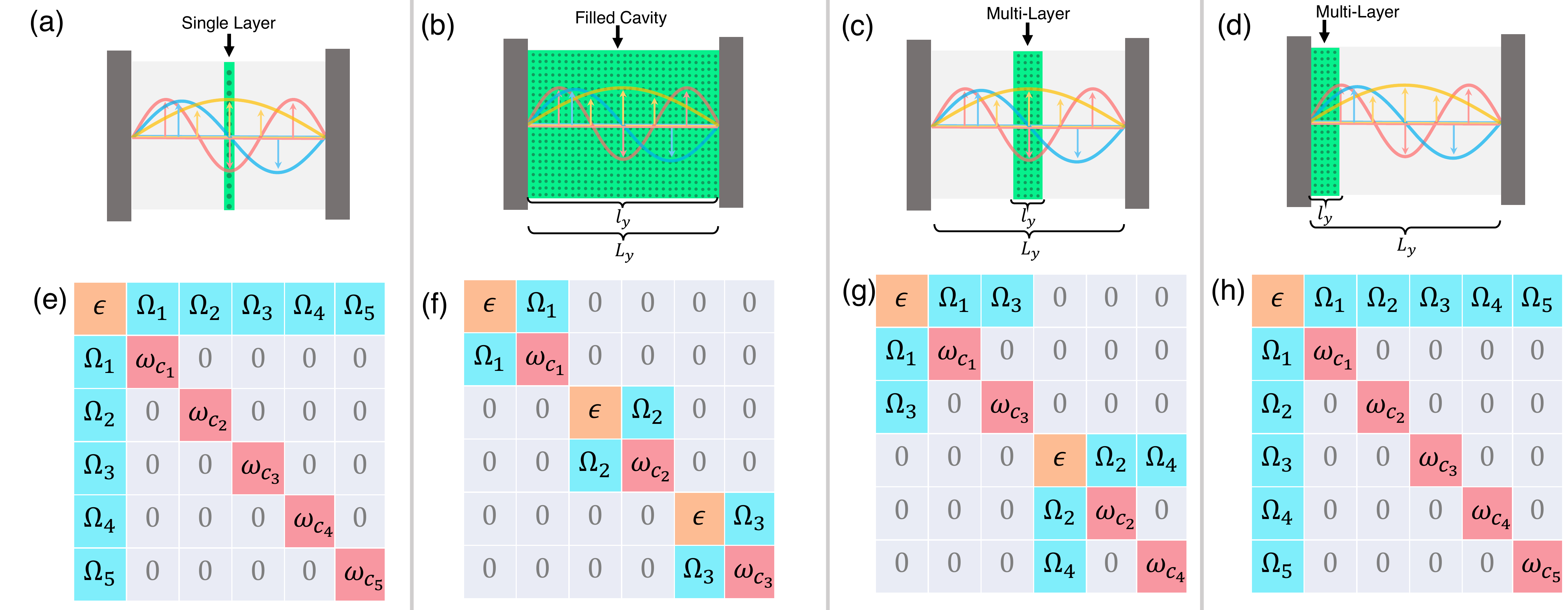}
\caption{\footnotesize  Schematic illustration of various cavity setups in (a)-(d) and their corresponding light-matter matrix  in (e)-(h) that can be diagonalized to obtain multi-mode polariton dispersion. The solid green box represents the material of various thicknesses $l_y$ and centered at various locations, (a) a single layer material located near the center, (b) material thickness same as the length of the cavity, (c) multi-layer material thickness much smaller than energetically relevant cavity mode wavelengths $l_y \ll {2\pi}{k_y}$ located at the center of the cavity (d) same as c but located beside one of the cavity mirrors. }
\label{fig1}
\end{figure*}

Next, we perform a $QR$ decomposition of $\Lambda_{\mathrm{NO}}$ to obtain the orthonormal matrix $\Lambda_{\mathrm{O}}$ (corresponding to $Q$).   
Using $\Lambda_{\mathrm{O}}$, we define a unitary matrix $U_\mathrm{O}$ of dimension $N_y \times N_y$ such that  $[U_\mathrm{O}]_{m, j} = [\Lambda_\mathrm{O}]_{m,j}$ for $j \in \{ n_\mathrm{min}, n_\mathrm{min}+1, ..., n_\mathrm{min} + N_e\}$   and the rest of the matrix elements  are chosen such that $\sum_m [U_\mathrm{O}]_{m,j'} \cdot [U_\mathrm{O}]_{m,j} = \delta_{jj'}$. Using the unitary matrix $U_\mathrm{O}$,  we define new matter excitation operator as,
 
 \begin{align}
 \hat{d}_{k_x,k_y} &= \sum_{m}[U_\mathrm{O}]_{m, n_y} \cdot \hat{d}_{k_x,m}, 
 \end{align}
where the index $k_y  = \frac{\pi}{L_y}n_y$. The consequence of this transformation is that for $N$ energetically relevant cavity modes, there are    $N_e \le N$ matter excitation operators $ \{\hat{d}_{k_x,k_y}\}$, such that $k_y \in \{n_\mathrm{min}\frac{\pi}{L_y}, ..., (n_\mathrm{min} + N_e)\frac{\pi}{L_y}\} \equiv \mathcal{N}_c$,  which couples to the photon operators $\{\hat{a}_{k_x,k_y}\}$. The rest of the matter excitation operators $\{\hat{d}_{k_x,k_y}\}$ for which $ k_y \notin \mathcal{N}_c$ are dark and can be dropped from the light-matter Hamiltonian.  The Hamiltonian ${\hat{h}}_{\mathrm{GTC}}(k_x)$ can be written  (using Eqn.~\ref{TC-k.x})  

\begin{align}\label{kx-GTC-0tau}
 {\hat{h}}_\mathrm{GTC}&(k_x)= \sum_{{  k_y  \in \mathcal{K}_c}} \hat{a}_{{k_x, k_y}}^{\dagger}\hat{a}_{{k_x, k_y}}\omega_c({k_x, k_y})+  \epsilon(k_x) \hat{d}^{\dagger}_{k_x, k_y}\hat{d}_{k_x, k_y} \nonumber\\
+  &\sum_{k_{y'}\in {\mathcal{N}_c}}\sum_{k_y \in {\mathcal{K}_c}}  \Omega({k_y,k_{y'}})  \big(  \hat{a}_{k_x, k_y}^\dagger \hat{d}_{ k_x, k_{y'}} +    \hat{a}_{{k_x, k_y}}\hat{d}_{ k_x, k_{y'}}^{\dagger} \big),
  \end{align}
where 
\begin{equation}
\Omega({k_y,k_{y'}}) = {g_{k_x,k_y}}     \sum_{m} \sin(k_y Y_m) \cdot [U_\mathrm{O}]_{m,n_{y'}}. 
\end{equation}
Here, 
$\mathcal{K}_c \equiv \{ n_\mathrm{min}\frac{\pi}{L_y}, (n_\mathrm{min}+1)\frac{\pi}{L_y},..., n_\mathrm{max}\frac{\pi}{L_y}  \}$
defines the subspace of energetically relevant cavity operators. Note the relation $\mathcal{N}_c \subset  \mathcal{K}_c$, where $\mathcal{N}_c$ is constructed such that columns of $\Lambda_\mathrm{NO}$ are linearly independent (in most cases $\mathcal{N}_c \equiv  \mathcal{K}_c$ and $N = N_e$).



A general  $(N+N_e)\times(N+N_e)$ matrix model when using the single excited subspace spanning $\{\hat{a}_{{k_x, k_y}}^\dagger| G,0\rangle,  \hat{d}_{{k_x, k_y}}^{^\dagger}|G,0\rangle \}$ (where $|G,0\rangle$ denote the ground state of matter with 0 cavity photons) is obtained as


\begin{align}\label{ML-general}
&{\hat{h}}_{\mathrm{N+N_e}}({k_x}) =  \\
&\left[\!\begin{array}{*{24}{c@{\hspace{2.4pt}}}}
\!\epsilon(k_x) & {\Omega}({\frac{\pi n_1}{L_y},\frac{\pi n_1}{L_y}})   & 0 & {\Omega}({\frac{\pi n_1}{L_y},\frac{\pi n_2}{L_y}}) & \hdots \\
 \!{\Omega}({\frac{\pi n_1}{L_y},\frac{\pi n_1}{L_y}})  & \omega_c({k_x, \frac{\pi n_1}{L_y}})    & {\Omega}({\frac{\pi n_2}{L_y},\frac{\pi n_1}{L_y}})  & 0  & \hdots \\
 \!0 & {\Omega}({\frac{\pi n_2}{L_y},\frac{\pi n_1}{L_y}})  & \epsilon(k_x) &  {\Omega}({\frac{\pi n_1}{L_y},\frac{\pi n_2}{L_y}})  & \hdots \\
 \!{\Omega}({\frac{\pi n_1}{L_y},\frac{\pi n_2}{L_y}}) & 0   & {\Omega}({\frac{\pi n_1}{L_y},\frac{\pi n_2}{L_y}}) & \omega_c({k_x, \frac{\pi n_2}{L_y}}) & \hdots \\
 \!\vdots    & \vdots    & \vdots & \vdots    & 
  \ddots\end{array}\!\right]. \nonumber
\end{align}

Note that when $N_e =1$, the above matrix reduces to the $(N+1)\times (N+1)$ form~\cite{Graf2016NC, Balasubrahmaniyam2021PRB} (such as for a single-layer material). Meanwhile, when $N_e = N$ and for ${\Omega}({\frac{\pi m}{L_y},\frac{\pi n}{L_y}}) = \delta_{m,n} {\Omega}({\frac{\pi n}{L_y},\frac{\pi n}{L_y}})$, the above matrix reduces to the $2N\times 2N$ (such as for a filled cavity) form~\cite{Richter2015APL,Georgiou2021JCP,Balasubrahmaniyam2021PRB}.

 In Fig.\ref{fig1} we briefly summarize the schematic forms of the Hamiltonian for various cavity setups that can be diagonalized to obtain the polariton dispersion. Using the general strategy outlined above, for a single layer  material shown in 
Fig.~\ref{fig1}e, we find the   widely used $(N+1)\times (N+1)$ matrix~\cite{ Dietrich2016SA, Coles2014APL, Michetti2005PRB, Richter2015APL,Georgiou2021JCP,Balasubrahmaniyam2021PRB,Orosz2011APE,Faure2009APL}. In this model, within each $k_x$ block, one exciton state, corresponding to $\hat{d}_{k_x,m=0}^{\dagger}|G,0\rangle$, is coupled to $N$ cavity excitations $\hat{a}_{k_x,k_y}^{\dagger}|G,0\rangle$ via the coupling $ {\Omega}_{n_y}  =    g_{\boldsymbol k} \sin( \frac{\pi n_y}{L_z} \cdot Y_0)$ where $Y_0$ is the position of the single layer inside cavity. 

For a filled cavity, we obtain an $2N\times 2N$ matrix that include $N$ exciton states $\hat{d}_{k_x,k_y}^{\dagger}|G,0\rangle$ where $\hat{d}_{k_x,k_y}^{\dagger} =  \frac{1}{\sqrt{\mathcal{N}_y}}\sin ( k_y \cdot (m-\frac{1}{2}) ) \hat{d}_{k_x,m}^{\dagger}$ with $ \mathcal{N}_y$ as a normalization constant, that are coupled to $N$ cavity excitations $\hat{a}_{k_x,k_y}^{\dagger}|G,0\rangle$. This $2N\times 2N$ matrix model contains $N$ non-interacting $2\times 2$ blocks as shown in Fig.~\ref{fig1}f.  Therefore, in a filled cavity there exists $N$ exciton that couple to  $N$ cavity modes of matching $k_y$. We also obtain the same  $2N\times 2N$ model when considering interacting layers $\tau_y \ne 0$ (see detailed analytical expressions in the SI).
 
For partially filled cavities we find two interesting  matrix models depending on the position of the material inside a cavity. For a thin material placed at the middle of the cavity we find a $(N+2)\times (N+2)$ matrix (Fig.~\ref{fig1}g) model which is composed of two isolated  $(\frac{N}{2}+1)\times (\frac{N}{2}+1)$ blocks. One block contains one `symmetric' exciton state $\hat{d}_{k_x,\mathrm{S}}^{\dagger}|G,0\rangle$ where $\hat{d}_{k_x,\mathrm{S}}^{\dagger} = \frac{1}{\sqrt{N_y}}  \sum_m  \hat{d}^{\dagger}_{k_x, m}$ is coupled to odd cavity modes ($k_y = \frac{n_y\pi }{L_z}$ with $n_y = 2,4,6 ...$). The other block contain one `asymmetric' exciton state $\hat{d}_{k_x,\mathrm{A}}^{\dagger}|G,0\rangle$ where $\hat{d}_{k_x,\mathrm{A}}^{\dagger} = \frac{1}{\sqrt{\mathcal{N}}}  \sum_m    \big(Y_m - \frac{L_y}{2}\big) \hat{d}^{\dagger}_{k_x, m}$ that is coupled to even 
cavity modes. For a thin material placed next to a mirror, we obtain the $(N+1)\times(N+1)$ model which is shown in Fig.~\ref{fig1}h. In this model one exciton state $\hat{d}_{k_x,B}^{\dagger}|G,0\rangle$ where $\hat{d}_{k_x, B} = {\frac{1}{\sqrt{\mathcal{N}_B}}} \sum_{m}Y_m \hat{d}_{k_x,m}$ ($\mathcal{N}_B$ is a normalization constant) is coupled to $N$ cavity modes. Note that despite this matrix model being structurally similar to the case of a single layer, the analytical forms of the couplings are quite different and are provided in the SI.

\begin{figure}[!t]
\centering
\includegraphics[width=1.0\linewidth]{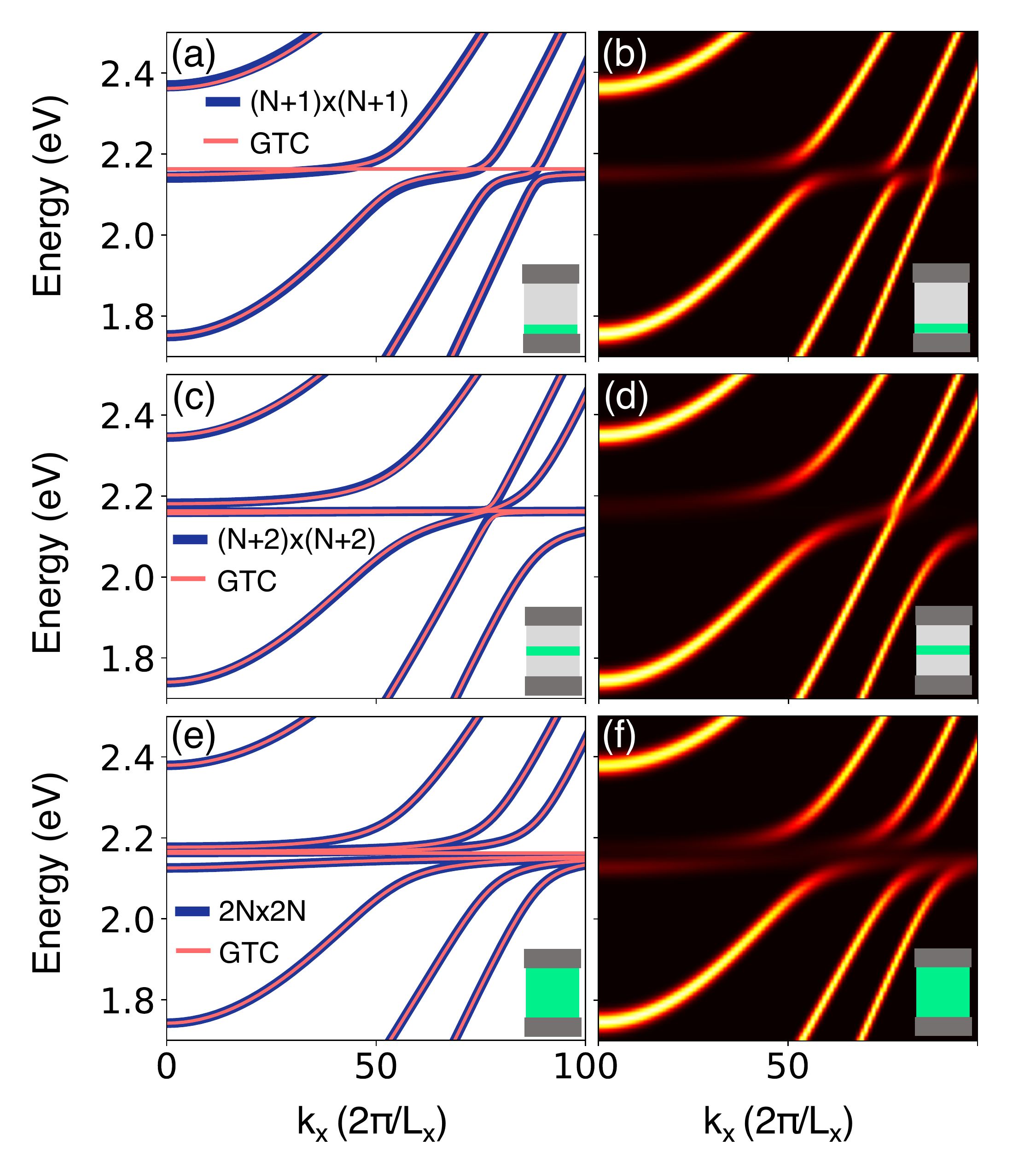}
\caption{\footnotesize Polariton dispersion for thin material placed (a)-(b) near a cavity mirror, (c)-(d) placed in the middle and a (e)-(f) filled cavity. (a), (c) and (e) Polariton dispersion computed from the generalized Tavis-Cummings Hamiltonian compared to the  $(N+1)\times (N+1)$ model in (a), $(N+2)\times (N+2)$ model in (c) and $2N\times 2N$ model in (e). Corresponding absorption spectra for the coupled molecule-cavity hybrid system in (b), (d) and (f).
}
\label{fig3}
\end{figure} 
    }

{\footnotesize

{\bf Numerical Results.} Fig.~\ref{fig3} presents numerical results for various multi-layered materials. In Fig.~\ref{fig3}a-d we consider a thin material which is either placed  next to a mirror (Fig.~\ref{fig3}a-b) or in the middle of the cavity (Fig.~\ref{fig3}c-d). We choose $L_y = 20000$ a.u. and a  thickness $l_y = \delta l_y \cdot N_y = 2000 \ll L_y$ with $\delta l_y =20$ a.u. and $N_y = 100$. The light-matter coupling used in Fig.~\ref{fig3}a-d is $g_c = 5$ meV, where $g_\mathrm{k} = \sqrt{\frac{\omega_c({k_x,k_y})}{\omega_c(0,\pi/L_y)}}g_c$,  and for the matter parameters we choose $\omega_0 = 2.2$ eV with $\tau_x = 150$ cm$^{-1}$.  The position of each layer of matter is given by  $Y_m = (m+\frac{1}{2}){\delta l_y}$. Here, we include five energetically relevant  cavity modes with $k_y \le 5\frac{\pi}{L_y}$. 

Fig.~\ref{fig3}a shows that the $(N+1)\times(N+1)$ model shown in Fig.~\ref{fig1}g (see details in Eqn.~S24 in the SI) provides visually identical results compared to direct diagonalization of $\mathcal{H}_\mathrm{GTC}(k_z = 0)$ given in Eqn.~\ref{TC-k.x}. Note that direct diagonalization also shows the dark matter states which we ignore while constructing  $(N+1)\times(N+1)$ model. These dark states do not show up in the absorption (visibility) spectrum as they do not have any photonic contributions. This can be observed in  Fig.~\ref{fig3}b, which presents the absorption spectrum~\cite{Lidzey2000S, Tichauer2021JCP} of the coupled cavity-matter system given by 

\begin{align}
I_\mathrm{A}(\omega, k_x) =   \sum_{i} |\langle P_{i,k_x}|\sum_{k_y} \hat{a}^{\dagger}_{k_x, k_y}|G,0\rangle|^2  e^{-\frac{(\mathcal{E}_{i,k_x} - E_G - \omega)^2}{2\Gamma_c^2}} ,
\end{align}
 where $| P_{i,k_x}\rangle $ is the ith polaritonic state that is an eigenstate of $\hat{h}_\mathrm{GTC}(k_x)$ with energy $\mathcal{E}_{i,k_x}$, and $E_G$ is the   energy of $|G,0\rangle$. Here $\Gamma_c = 15$ meV is a broadening parameter that is chosen to account for various sources of dissipation phenomenologically, such as cavity loss. 

 In Fig.~\ref{fig3}c-d we consider a material placed in the middle of the cavity such that $Y_m = \frac{L_y}{2} - \frac{ l_y}{2} + m\cdot \delta l_y$. Fig.~\ref{fig3}c show that the $({N}+2)\times ( {N}+2)$ model shown in Fig.~\ref{fig1}f (with details provided in Eqns.~S30-S31) correctly predicts the polariton bands in comparison to the numerical results obtained by directly diagonalizing $\hat{h}_\mathrm{GTC}({k_x})$. Note that in Fig.~\ref{fig3}a-b the anti-crossings grow gradually larger with respect to the $k_y$ of the cavity mode (which scales as $k_y^{3/2}$ as shown in the SI). By contrast, Fig.~\ref{fig3}c-d shows that the anti-crossings are large for odd $k_y$ while they are negligible for even $k_y$, as the spatial dependence for even cavity modes becomes zero at the center of the cavity. Thus, the location of the material inside the cavity plays a crucial role in setting the polariton dispersion.

 Fig.~\ref{fig3}e-f presents the dispersion  and absorption spectra for a filled cavity. Here, we use $g_c = 2$ meV and use $N_y = 999$ with $Y_0 = \frac{\delta l_y}{2} = 10$  and the rest of the parameters are maintained as in Fig.~\ref{fig3}a-d.  Fig.~\ref{fig3}e presents the corresponding multimode polariton dispersion computed numerically by diagonalizing Eq.~\ref{TC-k.x} compared with the $2N\times 2N$ model in Eqn.~S19. The $2N\times 2N$ model reproduces the band dispersion, as expected. Overall, the results in  Fig.~\ref{fig3} demonstrates the validity of the approximate semi-analytical models which are suitable for specific position and properties of the multi-layered material inside the cavity. 
 
  \begin{figure}[!t]
\centering
\includegraphics[width=1.0\linewidth]{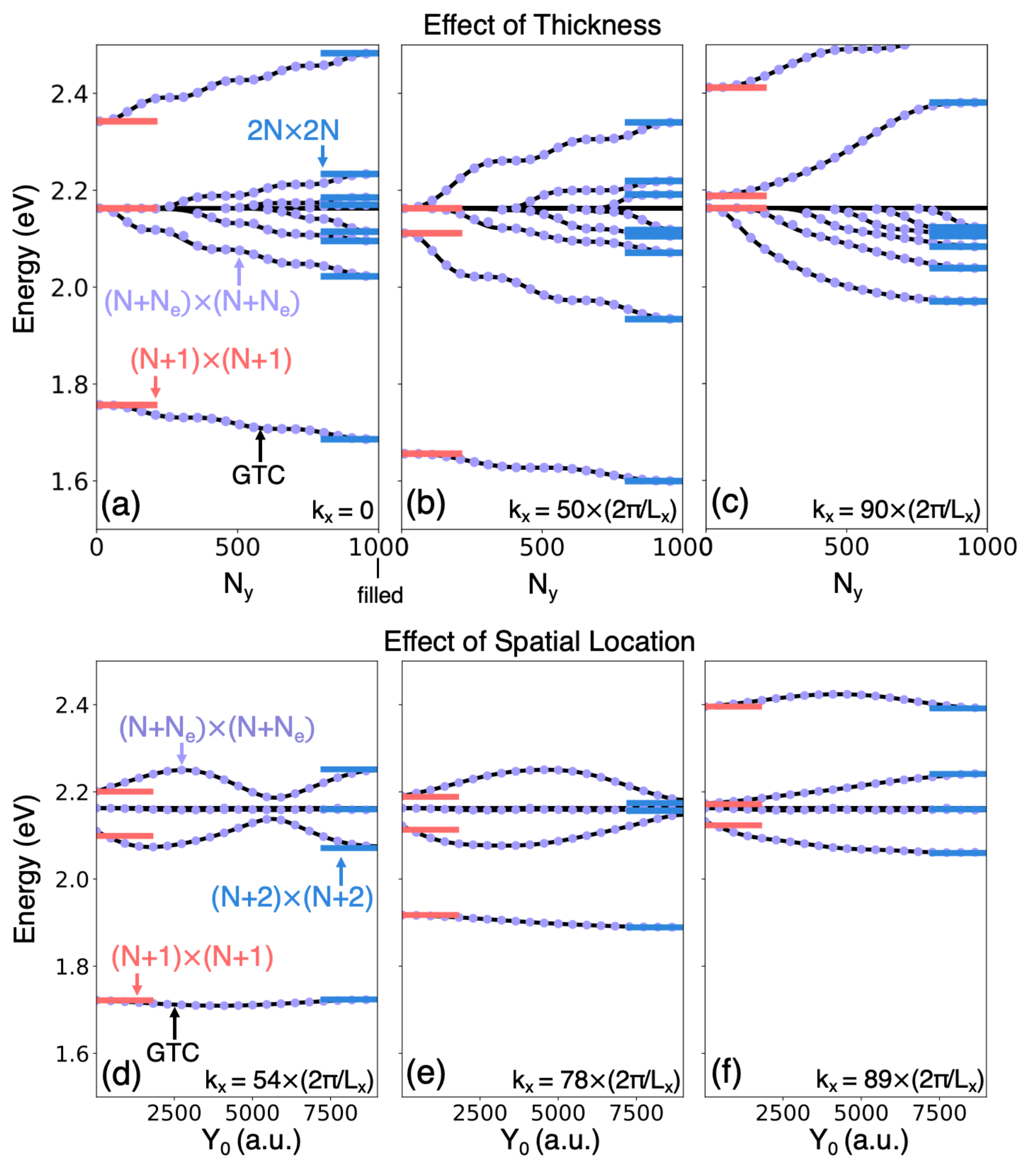}
\caption{\footnotesize   (a)-(c) Polariton dispersion as a function of the number of layers $N_y$ at various $k_x$ computed from generalized Tavis-Cummings (GTC) Hamiltonian compared to the predictions of the generalized $2N\times 2N$ (at $N_y = 999$) model and the $(N+1)\times (N+1)$ (at $N_y = 1$).  (d)-(f)  Polariton dispersion as a function of the position of the material inside cavity ($Y_0$ is the position of the first layer of the material) at various $k_x$ computed from generalized Tavis-Cummings (GTC) Hamiltonian compared to the predictions of $(N+1)\times (N+1)$ and $(N+2)\times (N+2)$ model. }
\label{fig5}
\end{figure} 

In  Fig.~\ref{fig5} we explore how the spatial location and the thickness of the material modifies the polariton dispersion. In Fig.~\ref{fig5}a-c we explore the dependence of the polariton dispersion on material thickness by plotting polariton bands as a function of $N_y$ (number of layers) at various $k_x$ values while keeping $g_c = 4.743$ meV a constant. Our numerical results show that the polariton dispersion at the two limiting extremes,  $N_y = 1$  (single layer) and $N_y = 999$ (filled), can be obtained using the $(N+1)\times (N+1)$ and the  $2N\times 2N$ models, respectively. For any intermediate values of $N_y$, {\it i.e.} partially filled cavities, accurate polariton dispersion can be obtained using the general   $(N+N_e)\times (N+N_e)$ model (compare the violet dots with the black solid lines) given in Eqn.~\ref{ML-general}.  
\begin{figure*}[!t]
\centering
\includegraphics[width=0.8\linewidth]{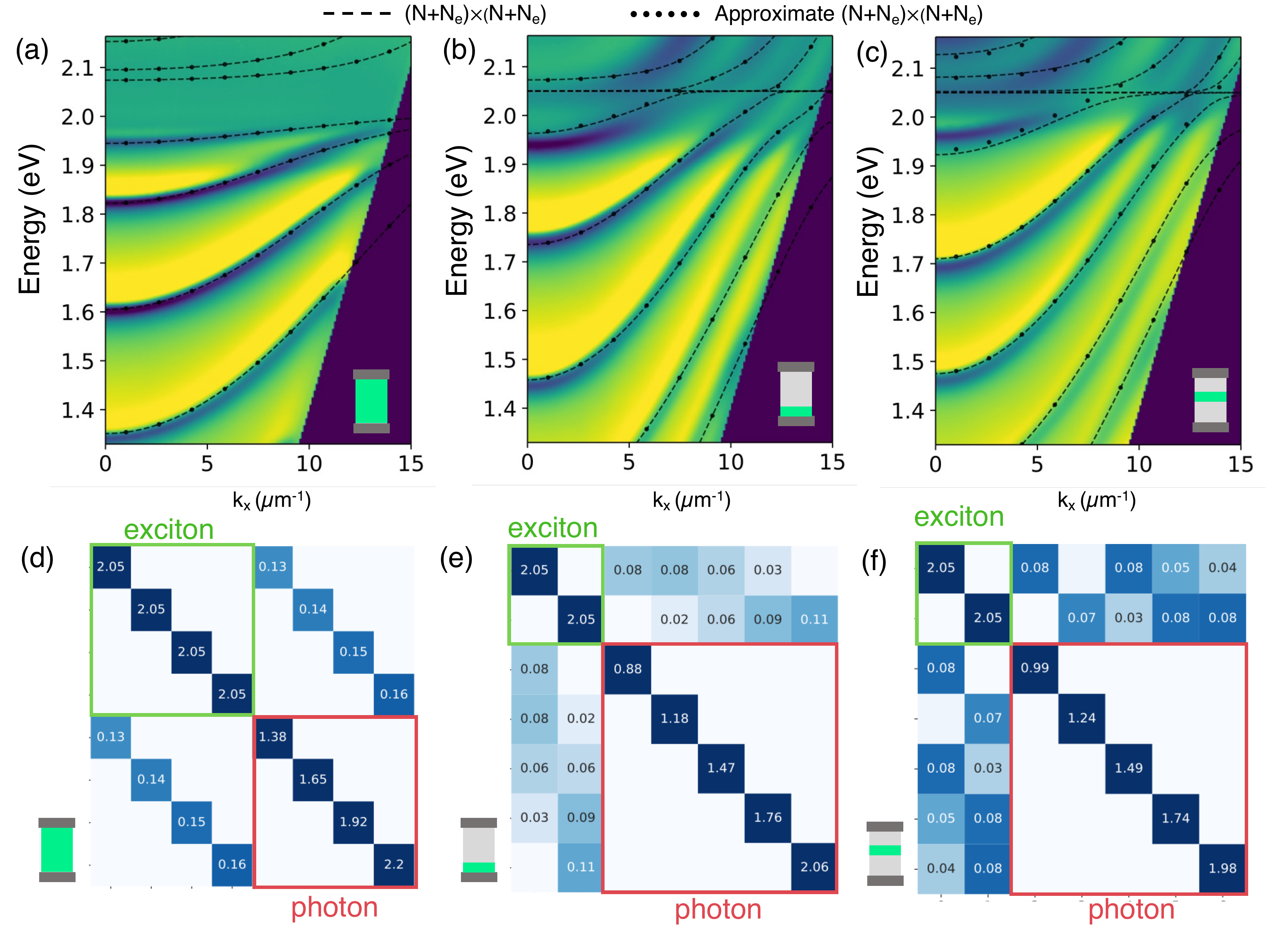}
\caption{\footnotesize (a)-(c)  Comparing experimental dispersion for three different structure (a) filled cavity, (b) material placed at the beside a mirror and (c) material placed near the middle of the cavity. The experimental results are compared to theoretical predictions with dashed line represents the predictions using the general  $(N+N_e)\times (N+N_e)$  model and the dotted lines represents the predictions using the approximate $(N+N_e)\times (N+N_e)$ model. (d)-(e) The matrix structure of the approximate $(N+N_e)\times (N+N_e)$ model corresponding to (a)-(c) with values in eV.}
\label{fig6}
\end{figure*}  
Similarly, Fig.~\ref{fig5}d-f presents the polariton bands at three specific value of $k_x$ and as a function of the material location $Y_0$. Here we use $g_c = 4.743$ meV while the rest of parameters are the same as in Fig.~\ref{fig3}a. The numerical results show that the in the limiting scenarios, at $Y_0 = \frac{\delta l_y}{2} = 10$ a.u. and $Y_0 = \frac{L_y - l_y}{2} = 9000$ a.u., the polariton dispersion can be obtained by the $(N+2)\times (N+2)$ and by $({N}+1)\times ({N}+1)$ models, respectively.

 Overall, we numerically demonstrate that the general   $(N+N_e)\times (N+N_e)$ model can be used to compute the multimode polariton dispersion for arbitrary geometries of multi-layered materials. We demonstrated that in specific scenarios, simplified models such as $2N\times 2N$, $(N+1)\times (N+1)$ and $(N+2)\times (N+2)$ model can be used to compute the polariton dispersion. Our work sheds light on the recently discussed ambiguity~\cite{Georgiou2021JCP,Balasubrahmaniyam2021PRB,Richter2015APL,Georgiou2021AC,Menghrajani2020ACSP,Tagami2021OE} of using $(N+1)\times (N+1)$   and $2N\times 2N$ models for fitting experimental dispersion. While for single-layer thin materials the $(N+1)\times (N+1)$ model is appropriate, the $2N\times 2N$ model   is  appropriate   for obtaining polariton dispersions in filled cavities. \textit{However, our result suggest that the general  $(N+N_e)\times (N+N_e)$ model should be used in all circumstances for extracting parameters related to light-matter coupling.}
 
{\bf Comparison with Experiment.} Here we implement the general   $(N+N_e)\times (N+N_e)$ model to compute the multimode polariton dispersion for a multi-layered perovskite material and compare 
to the experimentally obtained reflectance spectrum. We prepare a multi-layered 2D perovskite material BA$_2$(MA)$_2$Pb$_3$I$_{10}$, where BA = CH$_3$(CH$_2$)$_3$NH$_3$ and MA = CH$_3$NH$_3$ that is sandwiched between an Ag and Au layer that act as mirrors. We also use a Poly(methyl methacrylate)
 (PMMA) layer as a spacer to extend the quantization length of the cavity when making partially filled cavities. Experimental details can be found in the SI. Note that the layers of the perovskite material can be considered non-interacting, thus allowing us to directly implement the general  $(N+N_e)\times (N+N_e)$ model to obtain the multimode polariton dispersion.  


Fig.~\ref{fig6} presents the multimode polariton dispersion obtained experimentally via the reflectance spectra. Fig.~\ref{fig6}a shows the reflectance spectra obtained for filled cavity. To obtain the theoretical multimode polariton dispersion we  model the matter as a dispersionless material with a matter excitation energy of $\omega_0 = 2.05$ eV (such that $\epsilon(k_x) = \omega_0$) and a refractive index of $\eta_\mathrm{PX} = 2.2$.~\cite{SongACSML2021} We directly use this refractive index to obtain the uncoupled photon band dispersion $\omega_\mathrm{c} ({\boldsymbol{k}}) = \frac{c}{\eta} |\boldsymbol{k}|$ for the filled cavity presented in Fig.~\ref{fig6}a. We use an individual layer thickness of $\delta l_y = 50$ a.u.~\cite{MaoJACS2018} Note that for $\delta l_y \ll  l_y$  (that is for large $N_y$), the explicit value of the $\delta l_y$ does not matter. We find $ g_c \approx 4.2$ meV (such that $g_\mathrm{k} = \sqrt{\frac{\omega_c({k_x,k_y})}{\omega_c(0,\pi/L_y)}}g_c$) and $L_y \approx 19358$ a.u. through fitting and consider the 4th to 7th cavity mode branch along $y$, that is $k_y \in \{\frac{4\pi}{L_y}, \frac{5\pi}{L_y}, \frac{6\pi}{L_y}, \frac{7\pi}{L_y}\}$, for constructing the general $(N+N_e)\times (N+N_e)$ matrix. In addition to this we can also construct approximate $(N+N_e)\times (N+N_e)$ models by ignoring weakly coupled excitonic branches. A numerical way to perform this approximation is by using a tolerance factor $\epsilon_\mathrm{tol}$ (set to 0.1 here) below which the singular values are considered to be 0 when checking for linear independence via the use of Eqn.~\ref{NO-mat} for constructing $\Lambda_\mathrm{NO}$. The dispersion obtained with the general $(N+N_e)\times (N+N_e)$ model (dashed lines) and the approximate $(N+N_e)\times (N+N_e)$ model (black circles) is overlayed on the experimentally obtained reflectance in Fig.~\ref{fig6}a-c. These results show that our theoretical model is able to fit the experimental dispersion very well.  Note that the upper polariton branches are not visible in the experimental dispersion for this structure because the thick semiconductor absorbs too much above-gap light~\cite{FieramoscaSA2019,WangACSN2018}.


In Fig.~\ref{fig6}a only two parameters $g_c$ and $L_y$ are used for fitting the experimental dispersion. As can be seen, these parameters provide accurate band dispersion compared to the experimental results.  The numerically computed approximate $(N+N_e)\times (N+N_e)$ matrix model at $k_x = 0$ is shown in Fig,~\ref{fig6}d, which exactly takes the form of the widely used $2N\times 2N$ model\cite{ Michetti2005PRB, Richter2015APL,Georgiou2021JCP,Balasubrahmaniyam2021PRB}.  

Next in Fig.~\ref{fig6}b we manufacture a partially filled cavity by using PMMA as a spacer. The refractive index of the PMMA is $\eta_\mathrm{PMMA}$ = 1.5. For a cavity filled with materials of different refractive index, we obtain an effective refractive index through fitting such that $1.5 < \eta  < 2.2$. Here we obtain, the length of the perovskite material $l_y = 5550$ a.u., the length of the cavity $L_y \approx 24142$ a.u., $g_c \approx 5.39$ meV and the effective refractive index $\eta$ = 1.65   by fitting the dispersion curves visually. Here we have considered five energetically relevant cavity mode branches (the 3rd to 7th cavity mode branch along $y$) with $k_y \in \{\frac{3\pi}{L_y}, \frac{4\pi}{L_y}, \frac{5\pi}{L_y}, \frac{6\pi}{L_y}, \frac{7\pi}{L_y}\}$. These parameters fit the dispersion with reasonable accuracy. The numerically computed approximate $(N+N_e)\times (N+N_e)$ matrix at $k_x = 0$ is shown in Fig,~\ref{fig6}e for the case shown in Fig,~\ref{fig6}b. As can be seen here, the matrix has a $(N+2)\times (N+2)$ structure with a very complex set of couplings between the cavity and exciton branches.  

Finally, in Fig.~\ref{fig6}c we show results after fabrication of partially filled cavities with the material placed near the center of the cavity. Here, we fit and obtain $L_y = $ 28278 a.u.,  $l_y = $ 4500 a.u.,  $\eta$ = 1.67, $g_c \approx 5.146$ meV  and additionally find the location of the material $Y_0 = 5.1 \times L_y$, by visually fitting the dispersion. Here we have considered five energetically relevant cavity mode branches (3rd to 7th cavity mode branch along $y$). The corresponding approximate $(N+N_e)\times (N+N_e)$ matrix at $k_x = 0$ is shown in Fig.~\ref{fig6}e for the case shown in Fig.~\ref{fig6}f which, similar to the previous case, has a $(N+2)\times (N+2)$ structure with a complex set of couplings. In all cases we capture all of the subtle features of the experimental dispersion with reasonable accuracy. 

{\bf Conclusions.} In this work we have presented a microscopic theory for obtaining multimode polariton dispersion of a material inside a cavity. Starting with the GTC Hamiltonian, we develop a general strategy to obtain the polariton dispersion in multimode cavities, which takes the form of a general $(N+N_e)\times (N+N_e)$ model. Unlike the widely used $2N\times 2N$ or $(N+1)\times (N+1)$ models, our approach can be used for a material of arbitrary thickness and position within the cavity to obtain the multimode polariton dispersion. In contrast to previous approaches of directly fitting matrix elements, our method relies on structural parameters like the thickness and location of the material inside the cavity. We show that in certain limiting scenarios, the general  $(N+N_e)\times (N+N_e)$ model reduces to the widely used $2N\times 2N$ (for a filled cavity) model or the $(N+1)\times (N+1)$ (single or thin layer placed beside a mirror) model, and find other interesting models such as the $(N+2)\times (N+2)$ model that describes a thin material placed in the middle of the cavity.

While in this work we have employed a simple tight-binding model for the matter, our strategy can be generalized to using {\it ab initio} electronic structure theory with multiple bands. In future research, we will extend this approach toward {\it ab initio} modeling. 

}

\section{Acknowledgements}

{\footnotesize
This work was supported by NSF-1954791 (A.M. and D.R.R.) and NSF CHE-2203844 (M.E.D. and D.X.). We acknowledge the services provided by XSEDE (TG-CHE210085) and the OSG Consortium~\cite{osg09,osg07}, supported by the NSF awards $\#$2030508 and $\#$1836650.  

}
 
{\footnotesize
\bibliography{bib.bib} 

\providecommand{\latin}[1]{#1}
\makeatletter
\providecommand{\doi}
  {\begingroup\let\do\@makeother\dospecials
  \catcode`\{=1 \catcode`\}=2 \doi@aux}
\providecommand{\doi@aux}[1]{\endgroup\texttt{#1}}
\makeatother
\providecommand*\mcitethebibliography{\thebibliography}
\csname @ifundefined\endcsname{endmcitethebibliography}
  {\let\endmcitethebibliography\endthebibliography}{}
\begin{mcitethebibliography}{50}
\providecommand*\natexlab[1]{#1}
\providecommand*\mciteSetBstSublistMode[1]{}
\providecommand*\mciteSetBstMaxWidthForm[2]{}
\providecommand*\mciteBstWouldAddEndPuncttrue
  {\def\EndOfBibitem{\unskip.}}
\providecommand*\mciteBstWouldAddEndPunctfalse
  {\let\EndOfBibitem\relax}
\providecommand*\mciteSetBstMidEndSepPunct[3]{}
\providecommand*\mciteSetBstSublistLabelBeginEnd[3]{}
\providecommand*\EndOfBibitem{}
\mciteSetBstSublistMode{f}
\mciteSetBstMaxWidthForm{subitem}{(\alph{mcitesubitemcount})}
\mciteSetBstSublistLabelBeginEnd
  {\mcitemaxwidthsubitemform\space}
  {\relax}
  {\relax}

\bibitem[Hutchison \latin{et~al.}(2012)Hutchison, Schwartz, Genet, Devaux, and
  Ebbesen]{Hutchison2012ACIE}
Hutchison,~J.~A.; Schwartz,~T.; Genet,~C.; Devaux,~E.; Ebbesen,~T.~W. Modifying
  Chemical Landscapes by Coupling to Vacuum Fields. \emph{Angew. Chem. Int.
  Ed.} \textbf{2012}, \emph{51}, 1592--1596\relax
\mciteBstWouldAddEndPuncttrue
\mciteSetBstMidEndSepPunct{\mcitedefaultmidpunct}
{\mcitedefaultendpunct}{\mcitedefaultseppunct}\relax
\EndOfBibitem
\bibitem[Feist \latin{et~al.}(2018)Feist, Galego, and
  Garcia-Vidal]{Feist2018AP}
Feist,~J.; Galego,~J.; Garcia-Vidal,~F.~J. Polaritonic Chemistry with Organic
  Molecules. \emph{{ACS} Photonics} \textbf{2018}, \emph{5}, 205--216\relax
\mciteBstWouldAddEndPuncttrue
\mciteSetBstMidEndSepPunct{\mcitedefaultmidpunct}
{\mcitedefaultendpunct}{\mcitedefaultseppunct}\relax
\EndOfBibitem
\bibitem[Thomas \latin{et~al.}(2019)Thomas, Lethuillier-Karl, Nagarajan,
  Vergauwe, George, Chervy, Shalabney, Devaux, Genet, Moran, and
  Ebbesen]{Thomas2019S}
Thomas,~A.; Lethuillier-Karl,~L.; Nagarajan,~K.; Vergauwe,~R. M.~A.;
  George,~J.; Chervy,~T.; Shalabney,~A.; Devaux,~E.; Genet,~C.; Moran,~J.
  \latin{et~al.}  Tilting a ground-state reactivity landscape by vibrational
  strong coupling. \emph{Science} \textbf{2019}, \emph{363}, 615--619\relax
\mciteBstWouldAddEndPuncttrue
\mciteSetBstMidEndSepPunct{\mcitedefaultmidpunct}
{\mcitedefaultendpunct}{\mcitedefaultseppunct}\relax
\EndOfBibitem
\bibitem[Garcia-Vidal \latin{et~al.}(2021)Garcia-Vidal, Ciuti, and
  Ebbesen]{GarciaVidal2021S}
Garcia-Vidal,~F.~J.; Ciuti,~C.; Ebbesen,~T.~W. Manipulating matter by strong
  coupling to vacuum fields. \emph{Science} \textbf{2021}, \emph{373}\relax
\mciteBstWouldAddEndPuncttrue
\mciteSetBstMidEndSepPunct{\mcitedefaultmidpunct}
{\mcitedefaultendpunct}{\mcitedefaultseppunct}\relax
\EndOfBibitem
\bibitem[Nagarajan \latin{et~al.}(2021)Nagarajan, Thomas, and
  Ebbesen]{Nagarajan2021JACS}
Nagarajan,~K.; Thomas,~A.; Ebbesen,~T.~W. Chemistry under Vibrational Strong
  Coupling. \emph{J. Am. Chem. Soc.} \textbf{2021}, \emph{143},
  16877--16889\relax
\mciteBstWouldAddEndPuncttrue
\mciteSetBstMidEndSepPunct{\mcitedefaultmidpunct}
{\mcitedefaultendpunct}{\mcitedefaultseppunct}\relax
\EndOfBibitem
\bibitem[Ribeiro \latin{et~al.}(2018)Ribeiro, Mart{\'{\i}}nez-Mart{\'{\i}}nez,
  Du, Gonzalez-Angulo, and Yuen-Zhou]{Ribeiro2018CS}
Ribeiro,~R.~F.; Mart{\'{\i}}nez-Mart{\'{\i}}nez,~L.~A.; Du,~M.;
  Gonzalez-Angulo,~J.~C.; Yuen-Zhou,~J. Polariton chemistry: controlling
  molecular dynamics with optical cavities. \emph{Chem. Sci.} \textbf{2018},
  \emph{9}, 6325--6339\relax
\mciteBstWouldAddEndPuncttrue
\mciteSetBstMidEndSepPunct{\mcitedefaultmidpunct}
{\mcitedefaultendpunct}{\mcitedefaultseppunct}\relax
\EndOfBibitem
\bibitem[Mandal \latin{et~al.}(2020)Mandal, Krauss, and Huo]{Mandal2020JPCB}
Mandal,~A.; Krauss,~T.~D.; Huo,~P. Polariton-Mediated Electron Transfer via
  Cavity Quantum Electrodynamics. \emph{J. Phys. Chem. B} \textbf{2020},
  \emph{124}, 6321--6340\relax
\mciteBstWouldAddEndPuncttrue
\mciteSetBstMidEndSepPunct{\mcitedefaultmidpunct}
{\mcitedefaultendpunct}{\mcitedefaultseppunct}\relax
\EndOfBibitem
\bibitem[Semenov and Nitzan(2019)Semenov, and Nitzan]{Semenov2019JCP}
Semenov,~A.; Nitzan,~A. Electron transfer in confined electromagnetic fields.
  \emph{J. Chem. Phys.} \textbf{2019}, \emph{150}, 174122\relax
\mciteBstWouldAddEndPuncttrue
\mciteSetBstMidEndSepPunct{\mcitedefaultmidpunct}
{\mcitedefaultendpunct}{\mcitedefaultseppunct}\relax
\EndOfBibitem
\bibitem[Wu \latin{et~al.}(2022)Wu, Sifain, Delpo, and Scholes]{Weijun2022JCP}
Wu,~W.; Sifain,~A.~E.; Delpo,~C.~A.; Scholes,~G.~D. Polariton enhanced free
  charge carrier generation in donor--acceptor cavity systems by a
  second-hybridization mechanism. \emph{J. Chem. Phys.} \textbf{2022},
  \emph{157}, 161102\relax
\mciteBstWouldAddEndPuncttrue
\mciteSetBstMidEndSepPunct{\mcitedefaultmidpunct}
{\mcitedefaultendpunct}{\mcitedefaultseppunct}\relax
\EndOfBibitem
\bibitem[Mandal \latin{et~al.}(2022)Mandal, Taylor, Weight, Koessler, Li, and
  Huo]{Mandal2022CR}
Mandal,~A.; Taylor,~M.; Weight,~B.; Koessler,~E.; Li,~X.; Huo,~P. Theoretical
  Advances in Polariton Chemistry and Molecular Cavity Quantum Electrodynamics.
  \emph{ChemRxiv} \textbf{2022}, \relax
\mciteBstWouldAddEndPunctfalse
\mciteSetBstMidEndSepPunct{\mcitedefaultmidpunct}
{}{\mcitedefaultseppunct}\relax
\EndOfBibitem
\bibitem[Li \latin{et~al.}(2021)Li, Cui, Subotnik, and Nitzan]{Li2021ARPC}
Li,~T.~E.; Cui,~B.; Subotnik,~J.~E.; Nitzan,~A. Molecular Polaritonics:
  Chemical Dynamics Under Strong Light{\textendash}Matter Coupling. \emph{Annu.
  Rev. Phys. Chem.} \textbf{2021}, \emph{73}\relax
\mciteBstWouldAddEndPuncttrue
\mciteSetBstMidEndSepPunct{\mcitedefaultmidpunct}
{\mcitedefaultendpunct}{\mcitedefaultseppunct}\relax
\EndOfBibitem
\bibitem[Berghuis \latin{et~al.}(2022)Berghuis, Tichauer, de~Jong, Sokolovskii,
  Bai, Ramezani, Murai, Groenhof, and Rivas]{Berghuis2022AP}
Berghuis,~A.~M.; Tichauer,~R.~H.; de~Jong,~L. M.~A.; Sokolovskii,~I.; Bai,~P.;
  Ramezani,~M.; Murai,~S.; Groenhof,~G.; Rivas,~J.~G. Controlling Exciton
  Propagation in Organic Crystals through Strong Coupling to Plasmonic
  Nanoparticle Arrays. \emph{ACS Photonics} \textbf{2022}, \emph{9},
  2263--2272\relax
\mciteBstWouldAddEndPuncttrue
\mciteSetBstMidEndSepPunct{\mcitedefaultmidpunct}
{\mcitedefaultendpunct}{\mcitedefaultseppunct}\relax
\EndOfBibitem
\bibitem[Xu \latin{et~al.}(2022)Xu, Mandal, Baxter, Cheng, Lee, Su, Liu,
  Reichman, and Delor]{Xu2022}
Xu,~D.; Mandal,~A.; Baxter,~J.~M.; Cheng,~S.-W.; Lee,~I.; Su,~H.; Liu,~S.;
  Reichman,~D.~R.; Delor,~M. Ultrafast imaging of coherent polariton
  propagation and interactions. \emph{arXiv} \textbf{2022}, \relax
\mciteBstWouldAddEndPunctfalse
\mciteSetBstMidEndSepPunct{\mcitedefaultmidpunct}
{}{\mcitedefaultseppunct}\relax
\EndOfBibitem
\bibitem[Deng \latin{et~al.}(2010)Deng, Haug, and Yamamoto]{Deng2010RMP}
Deng,~H.; Haug,~H.; Yamamoto,~Y. Exciton-polariton Bose-Einstein condensation.
  \emph{Rev. Mod. Phys.} \textbf{2010}, \emph{82}, 1489--1537\relax
\mciteBstWouldAddEndPuncttrue
\mciteSetBstMidEndSepPunct{\mcitedefaultmidpunct}
{\mcitedefaultendpunct}{\mcitedefaultseppunct}\relax
\EndOfBibitem
\bibitem[Kockum \latin{et~al.}(2019)Kockum, Miranowicz, Liberato, Savasta, and
  Nori]{Kockum2019NRP}
Kockum,~A.~F.; Miranowicz,~A.; Liberato,~S.~D.; Savasta,~S.; Nori,~F.
  Ultrastrong coupling between light and matter. \emph{Nat. Rev. Phys.}
  \textbf{2019}, \emph{1}, 19--40\relax
\mciteBstWouldAddEndPuncttrue
\mciteSetBstMidEndSepPunct{\mcitedefaultmidpunct}
{\mcitedefaultendpunct}{\mcitedefaultseppunct}\relax
\EndOfBibitem
\bibitem[Keeling and K\'{e}na-Cohen(2020)Keeling, and
  K\'{e}na-Cohen]{Keeling2020ARPC}
Keeling,~J.; K\'{e}na-Cohen,~S. Bose–Einstein Condensation of
  Exciton-Polaritons in Organic Microcavities. \emph{Ann. Rev. Phys. Chem.}
  \textbf{2020}, \emph{71}, 435--459\relax
\mciteBstWouldAddEndPuncttrue
\mciteSetBstMidEndSepPunct{\mcitedefaultmidpunct}
{\mcitedefaultendpunct}{\mcitedefaultseppunct}\relax
\EndOfBibitem
\bibitem[Arnardottir \latin{et~al.}(2020)Arnardottir, Moilanen, Strashko,
  T{\"{o}}rm{\"{a}}, and Keeling]{Arnardottir2020PRL}
Arnardottir,~K.~B.; Moilanen,~A.~J.; Strashko,~A.; T{\"{o}}rm{\"{a}},~P.;
  Keeling,~J. Multimode Organic Polariton Lasing. \emph{Phys. Rev. Lett.}
  \textbf{2020}, \emph{125}, 233603\relax
\mciteBstWouldAddEndPuncttrue
\mciteSetBstMidEndSepPunct{\mcitedefaultmidpunct}
{\mcitedefaultendpunct}{\mcitedefaultseppunct}\relax
\EndOfBibitem
\bibitem[Rozenman \latin{et~al.}(2018)Rozenman, Akulov, Golombek, and
  Schwartz]{Rozenman2018ACSP}
Rozenman,~G.~G.; Akulov,~K.; Golombek,~A.; Schwartz,~T. Long-Range Transport of
  Organic Exciton-Polaritons Revealed by Ultrafast Microscopy. \emph{ACS
  Photonics} \textbf{2018}, \emph{5}, 105--110\relax
\mciteBstWouldAddEndPuncttrue
\mciteSetBstMidEndSepPunct{\mcitedefaultmidpunct}
{\mcitedefaultendpunct}{\mcitedefaultseppunct}\relax
\EndOfBibitem
\bibitem[Perez-Sanchez and Yuen-Zhou(2020)Perez-Sanchez, and
  Yuen-Zhou]{Sanchez2020JPCL}
Perez-Sanchez,~J.~B.; Yuen-Zhou,~J. Polariton Assisted Down-Conversion of
  Photons via Nonadiabatic Molecular Dynamics: A Molecular Dynamical Casimir
  Effect. \emph{J. Phys. Chem. Lett.} \textbf{2020}, \emph{11}, 152--159\relax
\mciteBstWouldAddEndPuncttrue
\mciteSetBstMidEndSepPunct{\mcitedefaultmidpunct}
{\mcitedefaultendpunct}{\mcitedefaultseppunct}\relax
\EndOfBibitem
\bibitem[Mandal \latin{et~al.}(2020)Mandal, Vega, and Huo]{Mandal2020JPCL}
Mandal,~A.; Vega,~S.~M.; Huo,~P. Polarized Fock States and the Dynamical
  Casimir Effect in Molecular Cavity Quantum Electrodynamics. \emph{J. Phys.
  Chem. Lett.} \textbf{2020}, \emph{11}, 9215--9223\relax
\mciteBstWouldAddEndPuncttrue
\mciteSetBstMidEndSepPunct{\mcitedefaultmidpunct}
{\mcitedefaultendpunct}{\mcitedefaultseppunct}\relax
\EndOfBibitem
\bibitem[Sanchez-Barquilla \latin{et~al.}(2022)Sanchez-Barquilla,
  Fernandez-Dominguez, Feist, and Garcia-Vidal]{Barquilla2022ACSP}
Sanchez-Barquilla,~M.; Fernandez-Dominguez,~A.~I.; Feist,~J.;
  Garcia-Vidal,~F.~J. A Theoretical Perspective on Molecular Polaritonics.
  \emph{ACS Photonics} \textbf{2022}, \emph{9}, 1830--1841\relax
\mciteBstWouldAddEndPuncttrue
\mciteSetBstMidEndSepPunct{\mcitedefaultmidpunct}
{\mcitedefaultendpunct}{\mcitedefaultseppunct}\relax
\EndOfBibitem
\bibitem[Reithmaier \latin{et~al.}(2004)Reithmaier, S{{e}}k, L{\"o}ffler,
  Hofmann, Kuhn, Reitzenstein, Keldysh, Kulakovskii, Reinecke, and
  Forchel]{Reithmaier2004N}
Reithmaier,~J.~P.; S{{e}}k,~G.; L{\"o}ffler,~A.; Hofmann,~C.; Kuhn,~S.;
  Reitzenstein,~S.; Keldysh,~L.; Kulakovskii,~V.; Reinecke,~T.; Forchel,~A.
  Strong coupling in a single quantum dot--semiconductor microcavity system.
  \emph{Nature} \textbf{2004}, \emph{432}, 197--200\relax
\mciteBstWouldAddEndPuncttrue
\mciteSetBstMidEndSepPunct{\mcitedefaultmidpunct}
{\mcitedefaultendpunct}{\mcitedefaultseppunct}\relax
\EndOfBibitem
\bibitem[M{\"{u}}ller \latin{et~al.}(2015)M{\"{u}}ller, Fischer, Rundquist,
  Dory, Lagoudakis, Sarmiento, Kelaita, Borish, and
  Vu{\v{c}}kovi{\'{c}}]{Muller2015PRX}
M{\"{u}}ller,~K.; Fischer,~K.~A.; Rundquist,~A.; Dory,~C.; Lagoudakis,~K.~G.;
  Sarmiento,~T.; Kelaita,~Y.~A.; Borish,~V.; Vu{\v{c}}kovi{\'{c}},~J. Ultrafast
  Polariton-Phonon Dynamics of Strongly Coupled Quantum Dot-Nanocavity Systems.
  \emph{Phys. Rev. X} \textbf{2015}, \emph{5}, 031006\relax
\mciteBstWouldAddEndPuncttrue
\mciteSetBstMidEndSepPunct{\mcitedefaultmidpunct}
{\mcitedefaultendpunct}{\mcitedefaultseppunct}\relax
\EndOfBibitem
\bibitem[Laussy \latin{et~al.}(2012)Laussy, del Valle, Schrapp, Laucht, and
  Finley]{Laussy2012JP}
Laussy,~F.~P.; del Valle,~E.; Schrapp,~M.; Laucht,~A.; Finley,~J.~J. Climbing
  the Jaynes-Cummings ladder by photon counting. \emph{J. Nanophotonics}
  \textbf{2012}, \emph{6}, 061803\relax
\mciteBstWouldAddEndPuncttrue
\mciteSetBstMidEndSepPunct{\mcitedefaultmidpunct}
{\mcitedefaultendpunct}{\mcitedefaultseppunct}\relax
\EndOfBibitem
\bibitem[Deng \latin{et~al.}(2003)Deng, Weihs, Snoke, Bloch, and
  Yamamoto]{Deng2003PNAS}
Deng,~H.; Weihs,~G.; Snoke,~D.; Bloch,~J.; Yamamoto,~Y. Polariton lasing vs.
  photon lasing in a semiconductor microcavity. \emph{Proc. Natl. Acad. Sci.
  U.S.A.} \textbf{2003}, \emph{100}, 15318--15323\relax
\mciteBstWouldAddEndPuncttrue
\mciteSetBstMidEndSepPunct{\mcitedefaultmidpunct}
{\mcitedefaultendpunct}{\mcitedefaultseppunct}\relax
\EndOfBibitem
\bibitem[Georgiou \latin{et~al.}(2021)Georgiou, McGhee, Jayaprakash, and
  Lidzey]{Georgiou2021JCP}
Georgiou,~K.; McGhee,~K.~E.; Jayaprakash,~R.; Lidzey,~D.~G. Observation of
  photon-mode decoupling in a strongly coupled multimode microcavity. \emph{J.
  Chem. Phys.} \textbf{2021}, \emph{154}, 124309\relax
\mciteBstWouldAddEndPuncttrue
\mciteSetBstMidEndSepPunct{\mcitedefaultmidpunct}
{\mcitedefaultendpunct}{\mcitedefaultseppunct}\relax
\EndOfBibitem
\bibitem[Richter \latin{et~al.}(2015)Richter, Michalsky, Fricke, Sturm, Franke,
  Grundmann, and Schmidt-Grund]{Richter2015APL}
Richter,~S.; Michalsky,~T.; Fricke,~L.; Sturm,~C.; Franke,~H.; Grundmann,~M.;
  Schmidt-Grund,~R. Maxwell consideration of polaritonic quasi-particle
  Hamiltonians in multi-level systems. \emph{App. Phys. Lett.} \textbf{2015},
  \emph{107}, 231104\relax
\mciteBstWouldAddEndPuncttrue
\mciteSetBstMidEndSepPunct{\mcitedefaultmidpunct}
{\mcitedefaultendpunct}{\mcitedefaultseppunct}\relax
\EndOfBibitem
\bibitem[Michetti and La~Rocca(2005)Michetti, and La~Rocca]{Michetti2005PRB}
Michetti,~P.; La~Rocca,~G. Polariton states in disordered organic
  microcavities. \emph{Physical Review B} \textbf{2005}, \emph{71},
  115320\relax
\mciteBstWouldAddEndPuncttrue
\mciteSetBstMidEndSepPunct{\mcitedefaultmidpunct}
{\mcitedefaultendpunct}{\mcitedefaultseppunct}\relax
\EndOfBibitem
\bibitem[Tichauer \latin{et~al.}(2021)Tichauer, Feist, and
  Groenhof]{Tichauer2021JCP}
Tichauer,~R.~H.; Feist,~J.; Groenhof,~G. Multi-scale dynamics simulations of
  molecular polaritons: The effect of multiple cavity modes on polariton
  relaxation. \emph{J. Chem. Phys.} \textbf{2021}, \emph{154}, 104112\relax
\mciteBstWouldAddEndPuncttrue
\mciteSetBstMidEndSepPunct{\mcitedefaultmidpunct}
{\mcitedefaultendpunct}{\mcitedefaultseppunct}\relax
\EndOfBibitem
\bibitem[Gerace and Andreani(2007)Gerace, and Andreani]{Gerace2007PRB}
Gerace,~D.; Andreani,~L.~C. Quantum theory of exciton-photon coupling in
  photonic crystal slabs with embedded quantum wells. \emph{Physical Review B}
  \textbf{2007}, \emph{75}, 235325\relax
\mciteBstWouldAddEndPuncttrue
\mciteSetBstMidEndSepPunct{\mcitedefaultmidpunct}
{\mcitedefaultendpunct}{\mcitedefaultseppunct}\relax
\EndOfBibitem
\bibitem[Balasubrahmaniyam \latin{et~al.}(2021)Balasubrahmaniyam, Genet, and
  Schwartz]{Balasubrahmaniyam2021PRB}
Balasubrahmaniyam,~M.; Genet,~C.; Schwartz,~T. Coupling and decoupling of
  polaritonic states in multimode cavities. \emph{Physical Review B}
  \textbf{2021}, \emph{103}, L241407\relax
\mciteBstWouldAddEndPuncttrue
\mciteSetBstMidEndSepPunct{\mcitedefaultmidpunct}
{\mcitedefaultendpunct}{\mcitedefaultseppunct}\relax
\EndOfBibitem
\bibitem[Keeling(2012)]{Keeling2012}
Keeling,~J. \emph{Light-Matter Interactions and Quantum Optics}; University of
  St. Andrews, 2012\relax
\mciteBstWouldAddEndPuncttrue
\mciteSetBstMidEndSepPunct{\mcitedefaultmidpunct}
{\mcitedefaultendpunct}{\mcitedefaultseppunct}\relax
\EndOfBibitem
\bibitem[Li \latin{et~al.}(2020)Li, Golez, Mazza, Millis, Georges, and
  Eckstein]{JiajunPRB2020}
Li,~J.; Golez,~D.; Mazza,~G.; Millis,~A.~J.; Georges,~A.; Eckstein,~M.
  Electromagnetic coupling in tight-binding models for strongly correlated
  light and matter. \emph{Phys. Rev. B} \textbf{2020}, \emph{101}, 205140\relax
\mciteBstWouldAddEndPuncttrue
\mciteSetBstMidEndSepPunct{\mcitedefaultmidpunct}
{\mcitedefaultendpunct}{\mcitedefaultseppunct}\relax
\EndOfBibitem
\bibitem[Dmytruk and Schir\'o(2021)Dmytruk, and Schir\'o]{DmytrukPRB2021}
Dmytruk,~O.; Schir\'o,~M. Gauge fixing for strongly correlated electrons
  coupled to quantum light. \emph{Phys. Rev. B} \textbf{2021}, \emph{103},
  075131\relax
\mciteBstWouldAddEndPuncttrue
\mciteSetBstMidEndSepPunct{\mcitedefaultmidpunct}
{\mcitedefaultendpunct}{\mcitedefaultseppunct}\relax
\EndOfBibitem
\bibitem[Graf \latin{et~al.}(2016)Graf, Tropf, Zakharko, Zaumseil, and
  Gather]{Graf2016NC}
Graf,~A.; Tropf,~L.; Zakharko,~Y.; Zaumseil,~J.; Gather,~M.~C. Near-infrared
  exciton-polaritons in strongly coupled single-walled carbon nanotube
  microcavities. \emph{Nature Communications} \textbf{2016}, \emph{7},
  13078\relax
\mciteBstWouldAddEndPuncttrue
\mciteSetBstMidEndSepPunct{\mcitedefaultmidpunct}
{\mcitedefaultendpunct}{\mcitedefaultseppunct}\relax
\EndOfBibitem
\bibitem[Dietrich \latin{et~al.}(2016)Dietrich, Steude, Tropf, Schubert,
  Kronenberg, Ostermann, H{\"o}fling, and Gather]{Dietrich2016SA}
Dietrich,~C.~P.; Steude,~A.; Tropf,~L.; Schubert,~M.; Kronenberg,~N.~M.;
  Ostermann,~K.; H{\"o}fling,~S.; Gather,~M.~C. An exciton-polariton laser
  based on biologically produced fluorescent protein. \emph{Science advances}
  \textbf{2016}, \emph{2}, e1600666\relax
\mciteBstWouldAddEndPuncttrue
\mciteSetBstMidEndSepPunct{\mcitedefaultmidpunct}
{\mcitedefaultendpunct}{\mcitedefaultseppunct}\relax
\EndOfBibitem
\bibitem[Coles and Lidzey(2014)Coles, and Lidzey]{Coles2014APL}
Coles,~D.~M.; Lidzey,~D.~G. A ladder of polariton branches formed by coupling
  an organic semiconductor exciton to a series of closely spaced cavity-photon
  modes. \emph{Applied Physics Letters} \textbf{2014}, \emph{104}, 191108\relax
\mciteBstWouldAddEndPuncttrue
\mciteSetBstMidEndSepPunct{\mcitedefaultmidpunct}
{\mcitedefaultendpunct}{\mcitedefaultseppunct}\relax
\EndOfBibitem
\bibitem[Orosz \latin{et~al.}(2011)Orosz, Reveret, Bouchoule,
  Z{\'u}{\~n}iga-P{\'e}rez, M{\'e}dard, Leymarie, Disseix, Mihailovic,
  Frayssinet, Semond, \latin{et~al.} others]{Orosz2011APE}
Orosz,~L.; Reveret,~F.; Bouchoule,~S.; Z{\'u}{\~n}iga-P{\'e}rez,~J.;
  M{\'e}dard,~F.; Leymarie,~J.; Disseix,~P.; Mihailovic,~M.; Frayssinet,~E.;
  Semond,~F. \latin{et~al.}  Fabrication and optical properties of a
  fully-hybrid epitaxial ZnO-based microcavity in the strong-coupling regime.
  \emph{Applied physics express} \textbf{2011}, \emph{4}, 072001\relax
\mciteBstWouldAddEndPuncttrue
\mciteSetBstMidEndSepPunct{\mcitedefaultmidpunct}
{\mcitedefaultendpunct}{\mcitedefaultseppunct}\relax
\EndOfBibitem
\bibitem[Faure \latin{et~al.}(2009)Faure, Brimont, Guillet, Bretagnon, Gil,
  M{\'e}dard, Lagarde, Disseix, Leymarie, Z{\'u}{\~n}iga-P{\'e}rez,
  \latin{et~al.} others]{Faure2009APL}
Faure,~S.; Brimont,~C.; Guillet,~T.; Bretagnon,~T.; Gil,~B.; M{\'e}dard,~F.;
  Lagarde,~D.; Disseix,~P.; Leymarie,~J.; Z{\'u}{\~n}iga-P{\'e}rez,~J.
  \latin{et~al.}  Relaxation and emission of Bragg-mode and cavity-mode
  polaritons in a ZnO microcavity at room temperature. \emph{Applied Physics
  Letters} \textbf{2009}, \emph{95}, 121102\relax
\mciteBstWouldAddEndPuncttrue
\mciteSetBstMidEndSepPunct{\mcitedefaultmidpunct}
{\mcitedefaultendpunct}{\mcitedefaultseppunct}\relax
\EndOfBibitem
\bibitem[Lidzey \latin{et~al.}(2000)Lidzey, Bradley, Armitage, Walker, and
  Skolnick]{Lidzey2000S}
Lidzey,~D.~G.; Bradley,~D. D.~C.; Armitage,~A.; Walker,~S.; Skolnick,~M.~S.
  Photon-Mediated Hybridization of Frenkel Excitons in Organic Semiconductor
  Microcavities. \emph{Science} \textbf{2000}, \emph{288}, 1620--1623\relax
\mciteBstWouldAddEndPuncttrue
\mciteSetBstMidEndSepPunct{\mcitedefaultmidpunct}
{\mcitedefaultendpunct}{\mcitedefaultseppunct}\relax
\EndOfBibitem
\bibitem[Georgiou \latin{et~al.}(2021)Georgiou, Jayaprakash, Othonos, and
  Lidzey]{Georgiou2021AC}
Georgiou,~K.; Jayaprakash,~R.; Othonos,~A.; Lidzey,~D.~G. Ultralong-Range
  Polariton-Assisted Energy Transfer in Organic Microcavities. \emph{Angewandte
  Chemie} \textbf{2021}, \emph{133}, 16797--16803\relax
\mciteBstWouldAddEndPuncttrue
\mciteSetBstMidEndSepPunct{\mcitedefaultmidpunct}
{\mcitedefaultendpunct}{\mcitedefaultseppunct}\relax
\EndOfBibitem
\bibitem[Menghrajani and Barnes(2020)Menghrajani, and
  Barnes]{Menghrajani2020ACSP}
Menghrajani,~K.~S.; Barnes,~W.~L. Strong Coupling beyond the Light-Line.
  \emph{ACS Photonics} \textbf{2020}, \emph{7}, 2448--2459, PMID:
  33163580\relax
\mciteBstWouldAddEndPuncttrue
\mciteSetBstMidEndSepPunct{\mcitedefaultmidpunct}
{\mcitedefaultendpunct}{\mcitedefaultseppunct}\relax
\EndOfBibitem
\bibitem[Tagami \latin{et~al.}(2021)Tagami, Ueda, Imai, Takahashi, Mizuno,
  Yanagi, Obuchi, Nakayama, and Yamashita]{Tagami2021OE}
Tagami,~T.; Ueda,~Y.; Imai,~K.; Takahashi,~S.; Mizuno,~H.; Yanagi,~H.;
  Obuchi,~T.; Nakayama,~M.; Yamashita,~K. Anisotropic light-matter coupling and
  below-threshold excitation dynamics in an organic crystal microcavity.
  \emph{Optics Express} \textbf{2021}, \emph{29}, 26433--26443\relax
\mciteBstWouldAddEndPuncttrue
\mciteSetBstMidEndSepPunct{\mcitedefaultmidpunct}
{\mcitedefaultendpunct}{\mcitedefaultseppunct}\relax
\EndOfBibitem
\bibitem[Song \latin{et~al.}(2021)Song, Hou, Wang, Sidhik, Miao, Gu, Zhang,
  Liu, Fakhraai, Even, Blancon, Mohite, and Jariwala]{SongACSML2021}
Song,~B.; Hou,~J.; Wang,~H.; Sidhik,~S.; Miao,~J.; Gu,~H.; Zhang,~H.; Liu,~S.;
  Fakhraai,~Z.; Even,~J. \latin{et~al.}  Determination of Dielectric Functions
  and Exciton Oscillator Strength of Two-Dimensional Hybrid Perovskites.
  \emph{ACS Materials Letters} \textbf{2021}, \emph{3}, 148--159\relax
\mciteBstWouldAddEndPuncttrue
\mciteSetBstMidEndSepPunct{\mcitedefaultmidpunct}
{\mcitedefaultendpunct}{\mcitedefaultseppunct}\relax
\EndOfBibitem
\bibitem[Mao \latin{et~al.}(2018)Mao, Stoumpos, and Kanatzidis]{MaoJACS2018}
Mao,~L.; Stoumpos,~C.~C.; Kanatzidis,~M.~G. Two-dimensional hybrid halide
  perovskites: principles and promises. \emph{J. Am. Chem. Soc.} \textbf{2018},
  \emph{141}, 1171--1190\relax
\mciteBstWouldAddEndPuncttrue
\mciteSetBstMidEndSepPunct{\mcitedefaultmidpunct}
{\mcitedefaultendpunct}{\mcitedefaultseppunct}\relax
\EndOfBibitem
\bibitem[Fieramosca \latin{et~al.}(2019)Fieramosca, Polimeno, Ardizzone, Marco,
  Pugliese, Maiorano, Giorgi, Dominici, Gigli, Gerace, Ballarini, and
  Sanvitto]{FieramoscaSA2019}
Fieramosca,~A.; Polimeno,~L.; Ardizzone,~V.; Marco,~L.~D.; Pugliese,~M.;
  Maiorano,~V.; Giorgi,~M.~D.; Dominici,~L.; Gigli,~G.; Gerace,~D.
  \latin{et~al.}  Two-dimensional hybrid perovskites sustaining strong
  polariton interactions at room temperature. \emph{Science Advances}
  \textbf{2019}, \emph{5}, eaav9967\relax
\mciteBstWouldAddEndPuncttrue
\mciteSetBstMidEndSepPunct{\mcitedefaultmidpunct}
{\mcitedefaultendpunct}{\mcitedefaultseppunct}\relax
\EndOfBibitem
\bibitem[Wang \latin{et~al.}(2018)Wang, Su, Xing, Bao, Diederichs, Liu, Liew,
  Chen, and Xiong]{WangACSN2018}
Wang,~J.; Su,~R.; Xing,~J.; Bao,~D.; Diederichs,~C.; Liu,~S.; Liew,~T.~C.;
  Chen,~Z.; Xiong,~Q. Room Temperature Coherently Coupled Exciton–Polaritons
  in Two-Dimensional Organic–Inorganic Perovskite. \emph{ACS Nano}
  \textbf{2018}, \emph{12}, 8382--8389\relax
\mciteBstWouldAddEndPuncttrue
\mciteSetBstMidEndSepPunct{\mcitedefaultmidpunct}
{\mcitedefaultendpunct}{\mcitedefaultseppunct}\relax
\EndOfBibitem
\bibitem[Sfiligoi \latin{et~al.}(2009)Sfiligoi, Bradley, Holzman, Mhashilkar,
  Padhi, and Wurthwein]{osg09}
Sfiligoi,~I.; Bradley,~D.~C.; Holzman,~B.; Mhashilkar,~P.; Padhi,~S.;
  Wurthwein,~F. The pilot way to grid resources using glideinWMS. 2009 WRI
  World Congress on Computer Science and Information Engineering. 2009; pp
  428--432\relax
\mciteBstWouldAddEndPuncttrue
\mciteSetBstMidEndSepPunct{\mcitedefaultmidpunct}
{\mcitedefaultendpunct}{\mcitedefaultseppunct}\relax
\EndOfBibitem
\bibitem[Pordes \latin{et~al.}(2007)Pordes, Petravick, Kramer, Olson, Livny,
  Roy, Avery, Blackburn, Wenaus, W{\"u}rthwein, Foster, Gardner, Wilde,
  Blatecky, McGee, and Quick]{osg07}
Pordes,~R.; Petravick,~D.; Kramer,~B.; Olson,~D.; Livny,~M.; Roy,~A.;
  Avery,~P.; Blackburn,~K.; Wenaus,~T.; W{\"u}rthwein,~F. \latin{et~al.}  The
  open science grid. J. Phys. Conf. Ser. 2007; p 012057\relax
\mciteBstWouldAddEndPuncttrue
\mciteSetBstMidEndSepPunct{\mcitedefaultmidpunct}
{\mcitedefaultendpunct}{\mcitedefaultseppunct}\relax
\EndOfBibitem
\end{mcitethebibliography}


\begin{thebibliography}{10}
\expandafter\ifx\csname url\endcsname\relax
  \def\url#1{\texttt{#1}}\fi
\expandafter\ifx\csname urlprefix\endcsname\relax\def\urlprefix{URL }\fi
\providecommand{\bibinfo}[2]{#2}
\providecommand{\eprint}[2][]{\url{#2}}

\bibitem{Steck}
\bibinfo{author}{Steck, D.~A.}
\newblock \bibinfo{title}{Quantum and atom optics}.
\newblock \emph{\bibinfo{journal}{available online at
  http://steck.us/teaching}}  (\bibinfo{year}{2018}).

\bibitem{Keeling2012}
\bibinfo{author}{Keeling, J.}
\newblock \emph{\bibinfo{title}{Light-Matter Interactions and Quantum Optics}}
  (\bibinfo{publisher}{University of St. Andrews}, \bibinfo{year}{2012}).

\bibitem{Mandal2022CR}
\bibinfo{author}{Mandal, A.} \emph{et~al.}
\newblock \bibinfo{title}{Theoretical advances in polariton chemistry and
  molecular cavity quantum electrodynamics}.
\newblock \emph{\bibinfo{journal}{ChemRxiv}}  (\bibinfo{year}{2022}).

\bibitem{JiajunPRB2020}
\bibinfo{author}{Li, J.} \emph{et~al.}
\newblock \bibinfo{title}{Electromagnetic coupling in tight-binding models for
  strongly correlated light and matter}.
\newblock \emph{\bibinfo{journal}{Phys. Rev. B}}
  \textbf{\bibinfo{volume}{101}}, \bibinfo{pages}{205140}
  (\bibinfo{year}{2020}).

\bibitem{DmytrukPRB2021}
\bibinfo{author}{Dmytruk, O.} \& \bibinfo{author}{Schir\'o, M.}
\newblock \bibinfo{title}{Gauge fixing for strongly correlated electrons
  coupled to quantum light}.
\newblock \emph{\bibinfo{journal}{Phys. Rev. B}}
  \textbf{\bibinfo{volume}{103}}, \bibinfo{pages}{075131}
  (\bibinfo{year}{2021}).

\bibitem{Combescot2008PRB}
\bibinfo{author}{Combescot, M.} \& \bibinfo{author}{Pogosov, W.}
\newblock \bibinfo{title}{Microscopic derivation of frenkel excitons in second
  quantization}.
\newblock \emph{\bibinfo{journal}{Physical Review B}}
  \textbf{\bibinfo{volume}{77}}, \bibinfo{pages}{085206}
  (\bibinfo{year}{2008}).

\bibitem{BassaniINCD1986}
\bibinfo{author}{Bassani, F.}, \bibinfo{author}{Ruggiero, F.} \&
  \bibinfo{author}{Quattropani, A.}
\newblock \bibinfo{title}{Microscopic quantum theory of exciton polaritons with
  spatial dispersion}.
\newblock \emph{\bibinfo{journal}{Il Nuovo Cimento D}}
  \textbf{\bibinfo{volume}{7}}, \bibinfo{pages}{700--716}
  (\bibinfo{year}{1986}).

\bibitem{Keeling2020ARPC}
\bibinfo{author}{Keeling, J.} \& \bibinfo{author}{K\'{e}na-Cohen, S.}
\newblock \bibinfo{title}{Bose–einstein condensation of exciton-polaritons in
  organic microcavities}.
\newblock \emph{\bibinfo{journal}{Ann. Rev. Phys. Chem.}}
  \textbf{\bibinfo{volume}{71}}, \bibinfo{pages}{435--459}
  (\bibinfo{year}{2020}).

\bibitem{Tichauer2021JCP}
\bibinfo{author}{Tichauer, R.~H.}, \bibinfo{author}{Feist, J.} \&
  \bibinfo{author}{Groenhof, G.}
\newblock \bibinfo{title}{Multi-scale dynamics simulations of molecular
  polaritons: The effect of multiple cavity modes on polariton relaxation}.
\newblock \emph{\bibinfo{journal}{J. Chem. Phys.}}
  \textbf{\bibinfo{volume}{154}}, \bibinfo{pages}{104112}
  (\bibinfo{year}{2021}).

\bibitem{Taylor2022OL}
\bibinfo{author}{Taylor, M. A.~D.}, \bibinfo{author}{Mandal, A.} \&
  \bibinfo{author}{Huo, P.}
\newblock \bibinfo{title}{Resolving ambiguities of the mode truncation in
  cavity quantum electrodynamics}.
\newblock \emph{\bibinfo{journal}{Opt. Lett.}} \textbf{\bibinfo{volume}{47}},
  \bibinfo{pages}{1446} (\bibinfo{year}{2022}).

\bibitem{Taylor2020PRL}
\bibinfo{author}{Taylor, M. A.~D.}, \bibinfo{author}{Mandal, A.},
  \bibinfo{author}{Zhou, W.} \& \bibinfo{author}{Huo, P.}
\newblock \bibinfo{title}{Resolution of gauge ambiguities in molecular cavity
  quantum electrodynamics}.
\newblock \emph{\bibinfo{journal}{Phys. Rev. Lett.}}
  \textbf{\bibinfo{volume}{125}}, \bibinfo{pages}{123602}
  (\bibinfo{year}{2020}).

\bibitem{Malekakhlagh2017PRL}
\bibinfo{author}{Malekakhlagh, M.}, \bibinfo{author}{Petrescu, A.} \&
  \bibinfo{author}{T\"ureci, H.~E.}
\newblock \bibinfo{title}{Cutoff-free circuit quantum electrodynamics}.
\newblock \emph{\bibinfo{journal}{Phys. Rev. Lett.}}
  \textbf{\bibinfo{volume}{119}}, \bibinfo{pages}{073601}
  (\bibinfo{year}{2017}).

\bibitem{Dietrich2016SA}
\bibinfo{author}{Dietrich, C.~P.} \emph{et~al.}
\newblock \bibinfo{title}{An exciton-polariton laser based on biologically
  produced fluorescent protein}.
\newblock \emph{\bibinfo{journal}{Science advances}}
  \textbf{\bibinfo{volume}{2}}, \bibinfo{pages}{e1600666}
  (\bibinfo{year}{2016}).

\bibitem{Coles2014APL}
\bibinfo{author}{Coles, D.~M.} \& \bibinfo{author}{Lidzey, D.~G.}
\newblock \bibinfo{title}{A ladder of polariton branches formed by coupling an
  organic semiconductor exciton to a series of closely spaced cavity-photon
  modes}.
\newblock \emph{\bibinfo{journal}{Applied Physics Letters}}
  \textbf{\bibinfo{volume}{104}}, \bibinfo{pages}{191108}
  (\bibinfo{year}{2014}).

\bibitem{Michetti2005PRB}
\bibinfo{author}{Michetti, P.} \& \bibinfo{author}{La~Rocca, G.}
\newblock \bibinfo{title}{Polariton states in disordered organic
  microcavities}.
\newblock \emph{\bibinfo{journal}{Physical Review B}}
  \textbf{\bibinfo{volume}{71}}, \bibinfo{pages}{115320}
  (\bibinfo{year}{2005}).

\bibitem{Richter2015APL}
\bibinfo{author}{Richter, S.} \emph{et~al.}
\newblock \bibinfo{title}{Maxwell consideration of polaritonic quasi-particle
  hamiltonians in multi-level systems}.
\newblock \emph{\bibinfo{journal}{App. Phys. Lett.}}
  \textbf{\bibinfo{volume}{107}}, \bibinfo{pages}{231104}
  (\bibinfo{year}{2015}).

\bibitem{Georgiou2021JCP}
\bibinfo{author}{Georgiou, K.}, \bibinfo{author}{McGhee, K.~E.},
  \bibinfo{author}{Jayaprakash, R.} \& \bibinfo{author}{Lidzey, D.~G.}
\newblock \bibinfo{title}{Observation of photon-mode decoupling in a strongly
  coupled multimode microcavity}.
\newblock \emph{\bibinfo{journal}{J. Chem. Phys.}}
  \textbf{\bibinfo{volume}{154}}, \bibinfo{pages}{124309}
  (\bibinfo{year}{2021}).

\bibitem{Balasubrahmaniyam2021PRB}
\bibinfo{author}{Balasubrahmaniyam, M.}, \bibinfo{author}{Genet, C.} \&
  \bibinfo{author}{Schwartz, T.}
\newblock \bibinfo{title}{Coupling and decoupling of polaritonic states in
  multimode cavities}.
\newblock \emph{\bibinfo{journal}{Physical Review B}}
  \textbf{\bibinfo{volume}{103}}, \bibinfo{pages}{L241407}
  (\bibinfo{year}{2021}).

\bibitem{Orosz2011APE}
\bibinfo{author}{Orosz, L.} \emph{et~al.}
\newblock \bibinfo{title}{Fabrication and optical properties of a fully-hybrid
  epitaxial zno-based microcavity in the strong-coupling regime}.
\newblock \emph{\bibinfo{journal}{Applied physics express}}
  \textbf{\bibinfo{volume}{4}}, \bibinfo{pages}{072001} (\bibinfo{year}{2011}).

\bibitem{Faure2009APL}
\bibinfo{author}{Faure, S.} \emph{et~al.}
\newblock \bibinfo{title}{Relaxation and emission of bragg-mode and cavity-mode
  polaritons in a zno microcavity at room temperature}.
\newblock \emph{\bibinfo{journal}{Applied Physics Letters}}
  \textbf{\bibinfo{volume}{95}}, \bibinfo{pages}{121102}
  (\bibinfo{year}{2009}).

\end{thebibliography}
 } 

\end{document}


{\footnotesize

\title{Supporting Information: Microscopic Theory of Multimode Polariton Dispersion in Multilayered Materials}

\author{ Arkajit Mandal$^1$\footnote{am5815@columbia.edu}~, Ding Xu$^1$, Ankit Mahajan$^1$, Joonho Lee$^1$, Milan E. Delor$^1$,  David R. Reichman$^1$\footnote{drr2103@columbia.edu}}

\maketitle

{ $^1$ Department of Chemistry, Columbia University, 3000 Broadway, New York, New York, 10027,  U.S.A}

\section{\normalsize Dipole Gauge Hamiltonian beyond Long-wavelength approximation}
Here we provide details for obtaining the dipole-gauge Hamiltonian used in the main text. The p.A Hamiltonian is given by (using atomic units $\hbar = 1$),
\begin{align}\label{eqn:HpA0}
\hat{H}_\mathrm{p \cdot A} &=   \sum_{u}  \frac{(\hat{\bf P}_u - e\hat{\bf A}(\hat{\bf Q}_u))^2}{2m_\mathrm{n}} +  \sum_{j} \frac{(\hat{\bf p}_j + e\hat{\bf A}(\hat{\bf r}_j))^2}{2m_e}   + V_\mathrm{coul}(\{{ {\bf r}_j, {\bf Q}_j}\}) + \sum_{{\boldsymbol k}, {\bf\xi}} \hat{a}_{{\boldsymbol k}, {\bf\xi}}^{\dagger}\hat{a}_{{\boldsymbol k}, {\boldsymbol{\xi}}}\omega_{\boldsymbol k},
\end{align}
where  $\hat{\bf  P}_j$ and $\hat{\bf  p}_j $  are the $j$th  canonical momentum operators with the corresponding nuclear position operator $\hat{\bf Q}_j$ and electronic position operator $\hat{\bf r}_j$ for nuclei and electrons. $\hat{\bf A}({\bf r})$ is the total vector potential at ${\bf r}$ and $\hat{a}_{{\boldsymbol k}, {\bf\xi}}^{\dagger}$ ($\hat{a}_{{\boldsymbol k}, {\bf\xi}}$) is the photon creation (annihilation) operator for radiation mode ${\bf k}$ with polarization ${\boldsymbol{\xi}} \in \{  {s}, { p} \}$. Note that we assume the charges of the nuclei are unity for simplicity, but the generalization beyond this is straightforward. The vector potential is then given as~\cite{Steck}
\begin{align}
&\hat{\bf A} ({\bf r} )  = \sum_{{\boldsymbol k}, {\boldsymbol \xi}} \hat{\bf A}_{{\boldsymbol k}, {\boldsymbol \xi}} ({\bf r} )
 = \sum_{{\boldsymbol k}, {\boldsymbol \xi}} \frac{{ \boldsymbol \lambda}_{{\boldsymbol k},   {\boldsymbol \xi}}}{\omega_c({\boldsymbol k})} \left[ e^{- i {\bf k_\parallel} \cdot {\bf r}} \hat{a}_{{\boldsymbol k},   {\boldsymbol \xi}}^\dagger +  e^{ i {\bf k_\parallel}  \cdot {\bf r}} \hat{a}_{{\boldsymbol k},   {\boldsymbol \xi}}\right]\sin({ k_y}   {  r_y} )  =  \sum_{{\boldsymbol k}, {\boldsymbol \xi}}\Big[ \mathcal{A}_{{\boldsymbol k},{\boldsymbol \xi}}({\bf r})\hat{a}_{{\boldsymbol k}}^{\dagger} +\mathcal{A}_{{\boldsymbol k},{\boldsymbol \xi}}^{*}({\bf r})\hat{a}_{{\boldsymbol k}}^{\dagger}\Big] \nonumber 
\end{align}
where ${ \boldsymbol \lambda}_{{\boldsymbol k}, {\boldsymbol \xi}} = \sqrt{\frac{\hbar \omega_c({\boldsymbol k})}{{\epsilon_0  \epsilon_r  \mathcal{V}}}} \hat{\boldsymbol e}_{{\boldsymbol k},   {\boldsymbol \xi}} $ with $\epsilon_0$  and $ \epsilon_r $ the vacuum and material permittivity, respectively, $\mathcal{V}$ is the quantization volume, $\hat{\boldsymbol e}_{{\boldsymbol k}, {\boldsymbol \xi}} \perp  {\boldsymbol k}$ is the polarization direction of the radiation mode ${\boldsymbol k} \equiv ( k_x, k_y, k_z)$ and ${\boldsymbol k_\parallel } = k_x \hat{\bf x} + k_z \hat{\bf z}$ is the longitudinal component of the cavity mode ${\boldsymbol k}$. 

First, we perform a nuclear-centered Power-Zienau-Woolley (PZW) transformation following Ref.~\cite{Keeling2012, Mandal2022CR} for the nuclear DOF with the unitary transformation operator $\hat{U}_\mathrm{nuc} = e^{- i \hat{\bf A}({\bf R}_u) \cdot \sum_u e\hat{\bf Q}_u }$ where ${\bf R}_u$ is the equilibrium position of $\hat{\bf Q}_u$ such that we can approximate 
$\hat{\bf A}(\hat{\bf Q}_u ) \approx \hat{\bf A}({\bf R}_u)$.  With this transformation we have,

\begin{align} 
&\hat{U}_\mathrm{nuc}^{\dagger}\hat{H}_\mathrm{p \cdot A} \hat{U}_\mathrm{nuc} =   \sum_{u}  \frac{\hat{\bf P}_u^2}{2m_\mathrm{n}} +  \sum_{j} \frac{(\hat{\bf p}_j + e\hat{\bf A}(\hat{\bf r}_j))^2}{2m_e}   + V_\mathrm{coul}(\{{ {\bf r}_j, {\bf Q}_j}\}) + \sum_{{\boldsymbol k}, {\bf\xi}} \Big(\hat{U}_\mathrm{nuc}^{\dagger} \hat{a}_{{\boldsymbol k}, {\bf\xi}}^{\dagger}\hat{U}_\mathrm{nuc} \Big)\Big(\hat{U}_\mathrm{nuc}^{\dagger}\hat{a}_{{\boldsymbol k}, {\boldsymbol{\xi}}}\omega_{\boldsymbol k}\hat{U}_\mathrm{nuc}\Big) \omega_{\boldsymbol k} \nonumber \\
&= \hat{ T}_{\bf Q} +  \sum_{j} \frac{(\hat{\bf p}_j + e\hat{\bf A}(\hat{\bf r}_j))^2}{2m_e}   + V_\mathrm{coul}(\{{ {\bf r}_j, {\bf Q}_j}\}) + \sum_{{\boldsymbol k}, {\bf\xi}} \Big( \hat{a}_{{\boldsymbol k}, {\bf\xi}}^{\dagger} + i \sum_u e\hat{\bf Q}_u\mathcal{A}^{*}_{{\boldsymbol k},{\boldsymbol \xi}} ({\bf R}_u)\Big)\Big( \hat{a}_{{\boldsymbol k}, {\boldsymbol{\xi}}} - i \sum_u e\hat{\bf Q}_u\mathcal{A}_{{\boldsymbol k},{\boldsymbol \xi}} ({\bf R}_u)\Big)\omega_{\boldsymbol k}~~~~.
\end{align}

Second, we perform an orbital centered PZW transformation following Ref.~\cite{JiajunPRB2020,DmytrukPRB2021} for the multi-electron system $\hat{U}_\mathrm{el} = e^{i \int d{\bf r} \Psi^{\dagger}({\bf r})  \chi({\bf r}) \Psi({\bf r}) }$ where $ \chi({\bf r}) = \int_{0}^{{\bf r}} \hat{\bf A} ({\bf s})\cdot {\bf ds}$, $\Psi^{\dagger}({\bf r}) = \sum_{u,\alpha}\Phi_{u,\alpha}({\bf r}) \hat{c}_{u,\alpha}^{\dagger}$ is the  electronic field operator with $\hat{c}_{u,\alpha}^{\dagger}$ as the spinless fermionic creation operator and $\{\Phi_{u,\alpha}({\bf r})\}$ are localized orthonormal single-particle wavefunctions that are centered around ${\bf R}_{u}$.  Note that  we  make a simplifying approximation of dropping of all spin degrees of
freedom  which also means that we drop all degeneracies coming from the orbital part of the electronic levels~\cite{Combescot2008PRB}. Overall, here  we consider a charge neutral system such that we have equal number of electron and charged nuclei. Using this transformation one obtains~\cite{DmytrukPRB2021} $\hat{H}^{'}_\mathrm{d \cdot E} = \hat{U}_\mathrm{el}^{\dagger}\hat{U}_\mathrm{nuc}^{\dagger}\hat{H}_\mathrm{p \cdot A} \hat{U}_\mathrm{nuc}  \hat{U}_\mathrm{el} $ as

\begin{align}
  \hat{H}^{'}_\mathrm{d \cdot E} 
 &= \hat{ T}_{\bf Q} + \hat{ T}_{\bf r}  + V_\mathrm{coul}(\{{ {\bf r}_j, {\bf Q}_j}\})  + \sum_{{\boldsymbol k}, {\bf\xi}} \Big( \hat{U}_\mathrm{el}^{\dagger} \hat{a}_{{\boldsymbol k}, {\bf\xi}}^{\dagger}\hat{U}_\mathrm{el} + i \sum_u e\hat{\bf Q}_u\mathcal{A}^{*}_{{\boldsymbol k},{\boldsymbol \xi}} ({\bf R}_u)\Big)\Big( \hat{U}_\mathrm{el}^{\dagger} \hat{a}_{{\boldsymbol k}, {\boldsymbol{\xi}}}\hat{U}_\mathrm{el}  - i \sum_u e\hat{\bf Q}_u\mathcal{A}_{{\boldsymbol k},{\boldsymbol \xi}} ({\bf R}_u)\Big)\omega_{\boldsymbol k}.  
\end{align}

Assuming that the spatial variation of $\mathcal{A}^{*}_{{\boldsymbol k},{\boldsymbol \xi}} ({\bf r})$ is negligible within the spatial extent of localized orbitals $\{\Phi_{u,\alpha}({\bf r})\}$   the following expression~\cite{DmytrukPRB2021} is obtained
\begin{align}
\hat{U}_\mathrm{el}^{\dagger} \hat{a}_{{\boldsymbol k}, {\bf\xi}}^{\dagger}\hat{U}_\mathrm{el} &= \hat{a}_{{\boldsymbol k}, {\bf\xi}}^{\dagger} -i\sum_{u,\alpha}e \Bigg[ \int_0^{{\bf R}_u}\mathcal{A}^{*}_{{\boldsymbol k},{\boldsymbol \xi}} (s)ds \Bigg] \hat{c}^{\dagger}_{u, \alpha} \hat{c}_{u, \alpha} - i \sum_{u,\alpha,\alpha'} e\mathcal{A}^{*}_{{\boldsymbol k},{\boldsymbol \xi}} ( {\bf R}_u) \Bigg[\int_{-\infty}^{\infty} d{\bf r}  \Phi_{u,\alpha}({\bf r}) ({\bf r} - {\bf R}_{u}) \Phi_{u,\alpha'}({\bf r})\Bigg] \hat{c}^{\dagger}_{u, \alpha} \hat{c}_{u, \alpha'}   \nonumber\\
&= \hat{a}_{{\boldsymbol k}, {\bf\xi}}^{\dagger} -i \sum_{u,\alpha} e\bar{\boldsymbol \chi}^{*}({\bf R}_u) \hat{c}^{\dagger}_{u, \alpha} \hat{c}_{u, \alpha} - i \sum_{u,\alpha,\alpha'} \mathcal{A}^{*}_{{\boldsymbol k},{\boldsymbol \xi}} ( {\bf R}_u){\boldsymbol \mu}_{\alpha,\alpha'} \hat{c}^{\dagger}_{u, \alpha} \hat{c}_{u, \alpha'}  
\end{align}
where we have defined $\int_0^{{\bf R}_u}\mathcal{A}^{*}_{{\boldsymbol k},{\boldsymbol \xi}} (s)ds = \bar{\boldsymbol \chi}^{*}({\bf R}_u)$ and the transition dipole moment ${\boldsymbol \mu}_{\alpha,\alpha'} = e\int_{-\infty}^{\infty} d{\bf r}  \Phi_{u,\alpha}({\bf r}) ({\bf r} - {\bf R}_{u}) \Phi_{u,\alpha'}({\bf r})$ and we set $\int_{-\infty}^{\infty} d{\bf r}  \Phi_{u,\alpha}({\bf r}) {\bf r} \Phi_{v,\alpha'}({\bf r}) = 0$ for $u \ne v$. Next we perform a polaron transformation with the unitary operator 
\begin{align}
\hat{U}_\mathrm{pol} = \exp\Big[i \sum_{{\boldsymbol k}, {\boldsymbol \xi}}\Big( \hat{a}_{{\boldsymbol k}, {\bf\xi}} \sum_{u,\alpha} e\bar{\boldsymbol \chi}^{*}({\bf R}_u) \hat{c}^{\dagger}_{u, \alpha} \hat{c}_{u, \alpha}+  \hat{a}_{{\boldsymbol k}, {\bf\xi}}^{\dagger} \sum_{u,\alpha} e\bar{\boldsymbol \chi}({\bf R}_u) \hat{c}^{\dagger}_{u, \alpha} \hat{c}_{u, \alpha} \Big) \Big]~~~.
\end{align}

Note the transformation on the fermionic creation operator which induces a Peierls phase~\cite{JiajunPRB2020}, 
\begin{align}
\hat{U}_\mathrm{pol}^{\dagger} \hat{c}_{u,\alpha} \hat{U}_\mathrm{pol} =   \hat{c}_{u,\alpha} \cdot \exp\Big[{i e\sum_{{\boldsymbol k}, {\boldsymbol \xi}} \hat{a}_{{\boldsymbol k}, {\bf\xi}}   \bar{\boldsymbol \chi}^{*}({\bf R}_u)  +  \hat{a}_{{\boldsymbol k}, {\bf\xi}}^{\dagger}  \bar{\boldsymbol \chi}({\bf R}_u)  }\Big]
\end{align}
while at the same time we can write 
$ \hat{U}_\mathrm{el}^{\dagger} \hat{a}_{{\boldsymbol k}, {\bf\xi}}^{\dagger}\hat{U}_\mathrm{el} = \hat{U}_\mathrm{pol} \hat{a}_{{\boldsymbol k}, {\bf\xi}}^{\dagger}  \hat{U}_\mathrm{pol}^{\dagger}  - i \sum_{u,\alpha,\alpha'} \mathcal{A}^{*}_{{\boldsymbol k},{\boldsymbol \xi}} ( {\bf R}_u){\boldsymbol \mu}_{\alpha,\alpha'} \hat{c}^{\dagger}_{u, \alpha} \hat{c}_{u, \alpha'} $. We write a polaron transformed d.E Hamiltonian~\cite{JiajunPRB2020} as

\begin{align}
\hat{U}_\mathrm{pol}^{\dagger}\hat{H}_\mathrm{d \cdot E} \hat{U}_\mathrm{pol} &= \hat{ T}_{\bf Q} + \hat{U}_\mathrm{pol}^{\dagger}(\hat{ T}_{\bf r}  + V_\mathrm{coul}(\{{ {\bf r}_j, {\bf Q}_j}\}) ) \hat{U}_\mathrm{pol} \nonumber \\
&+ \sum_{{\boldsymbol k}, {\bf\xi}} \Big(  \hat{a}_{{\boldsymbol k}, {\bf\xi}}^{\dagger}  + i \sum_u e\hat{\bf Q}_u\mathcal{A}^{*}_{{\boldsymbol k},{\boldsymbol \xi}} ({\bf R}_u)  - i \sum_{u,\alpha,\alpha'} \mathcal{A}^{*}_{{\boldsymbol k},{\boldsymbol \xi}} ( {\bf R}_u){\boldsymbol \mu}_{\alpha,\alpha'} \hat{c}^{\dagger}_{u, \alpha} \hat{c}_{u, \alpha'} \Big)\nonumber \\
&~~~\times\Big(  \hat{a}_{{\boldsymbol k}, {\boldsymbol{\xi}}}  - i \sum_u e\hat{\bf Q}_u\mathcal{A}_{{\boldsymbol k},{\boldsymbol \xi}} ({\bf R}_u) + i \sum_{u,\alpha,\alpha'} \mathcal{A}_{{\boldsymbol k},{\boldsymbol \xi}} ( {\bf R}_u){\boldsymbol \mu}_{\alpha,\alpha'} \hat{c}^{\dagger}_{u, \alpha} \hat{c}_{u, \alpha'}\Big)\omega_{\boldsymbol k}~~~~.
\end{align}

Next, we write $\mathcal{\hat{H}}_{el} = \hat{ T}_{\bf r}  + V_\mathrm{coul}(\{{ {\bf r}_j, {\bf Q}_j}\}) =   \sum_{\alpha , \alpha'} h_{u,v,\alpha,\alpha'} \hat{c}^{\dagger}_{u, \alpha} \hat{c}_{v, \alpha'} +  \sum_{\alpha,\alpha', \beta, \beta'}  \sum_{u,v,u',v'} U_{\alpha,\alpha', \beta, \beta'}^{u,v,u',v'} \hat{c}^{\dagger}_{u, \alpha} \hat{c}_{v, \alpha'}^{\dagger} \hat{c}_{u', \beta} \hat{c}_{v', \beta'} $ in the second quantization form. Note that
\begin{align}
\hat{U}_\mathrm{pol}^{\dagger}\hat{c}^{\dagger}_{u, \alpha} \hat{c}_{v, \alpha'} \hat{U}_\mathrm{pol} &= (\hat{U}_\mathrm{pol}^{\dagger}\hat{c}^{\dagger}_{u, \alpha}\hat{U}_\mathrm{pol} ) (\hat{U}_\mathrm{pol}^{\dagger} \hat{c}_{v, \alpha'} \hat{U}_\mathrm{pol}) \nonumber \\
&= \hat{c}^{\dagger}_{u, \alpha} \hat{c}_{v, \alpha'} \cdot  e^{i \sum_{{\boldsymbol k}, {\boldsymbol \xi}} \hat{a}_{{\boldsymbol k}, {\bf\xi}}   e (\bar{\boldsymbol \chi}^{*}({\bf R}_u) - \bar{\boldsymbol \chi}^{*}({\bf R}_v))  +  \hat{a}_{{\boldsymbol k}, {\bf\xi}}^{\dagger}  e(\bar{\boldsymbol \chi}({\bf R}_u) - \bar{\boldsymbol \chi}({\bf R}_v)  )} \nonumber\\
&\approx \hat{c}^{\dagger}_{u, \alpha} \hat{c}_{v, \alpha'} \cdot  e^{i e\sum_{{\boldsymbol k}, {\boldsymbol \xi}} \hat{a}_{{\boldsymbol k}, {\bf\xi}}   ({\bf R}_u \cdot \mathcal{A}^{*}_{{\boldsymbol k},{\boldsymbol \xi}} ( {\bf R}_u) - {\bf R}_v \cdot \mathcal{A}^{*}_{{\boldsymbol k},{\boldsymbol \xi}} ( {\bf R}_v))  +  \hat{a}_{{\boldsymbol k}, {\bf\xi}}^{\dagger}  ({\bf R}_u \cdot \mathcal{A}_{{\boldsymbol k},{\boldsymbol \xi}} ( {\bf R}_u) - {\bf R}_v \cdot \mathcal{A}_{{\boldsymbol k},{\boldsymbol \xi}} ( {\bf R}_v))} \nonumber\\
&=\hat{U}_\mathrm{pol}^{' \dagger}\hat{c}^{\dagger}_{u, \alpha} \hat{c}_{v, \alpha'} \hat{U}^{'}_\mathrm{pol} ,
\end{align}
where we have assumed that the spatial variation of the vector potential is negligible over spatial extent of  $ \Phi_{u,\alpha}({\bf r}) $ and $ \Phi_{v,\alpha}({\bf r}) $. Such a approximation will hold when $h_{u,v,\alpha,\alpha'}$ is only non-negligible when ${\bf R}_{u} \approx {\bf R}_{v}$, and   $U_{\alpha,\alpha', \beta, \beta'}^{u,v,u',v'}$ is also only non-negligible when ${\bf R}_{u} \approx {\bf R}_{u'}$  and ${\bf R}_{v} \approx {\bf R}_{v'}$.  Here we have defined $\hat{U}^{'}_\mathrm{pol} $ as
\begin{align}
\hat{U}'_\mathrm{pol} = \exp\Big[i \sum_{{\boldsymbol k}, {\boldsymbol \xi}}\Big( \hat{a}_{{\boldsymbol k}, {\bf\xi}} \sum_{u,\alpha} e {\bf R}_u \cdot \mathcal{A}^{*}_{{\boldsymbol k},{\boldsymbol \xi}} ( {\bf R}_u) \hat{c}^{\dagger}_{u, \alpha} \hat{c}_{u, \alpha}+  \hat{a}_{{\boldsymbol k}, {\bf\xi}}^{\dagger} \sum_{u,\alpha} e {\bf R}_u \cdot \mathcal{A}_{{\boldsymbol k},{\boldsymbol \xi}} ( {\bf R}_u) \hat{c}^{\dagger}_{u, \alpha} \hat{c}_{u, \alpha} \Big) \Big]~~~~.
\end{align}
We thus obtain the final form of the d.E Hamiltonian by performing the transformation with $U_{\phi} = \prod_{{\boldsymbol k},{\boldsymbol \xi}} e^{i\frac{\pi}{2}\hat{a}_{{\boldsymbol k},{\boldsymbol \xi}}^{\dagger}\hat{a}_{{\boldsymbol k},{\boldsymbol \xi}}}$ such that $U_{\phi}^{\dagger} \hat{a}_{{\boldsymbol k},{\boldsymbol \xi}} U_{\phi} = -i \hat{a}_{{\boldsymbol k},{\boldsymbol \xi}}$

\begin{align}
 \hat{H}_\mathrm{d \cdot E} &= U_{\phi}^{\dagger} \hat{H}_\mathrm{d \cdot E}^{'} U_{\phi} = U_{\phi}^{\dagger} (\hat{U}_\mathrm{pol}^{' } \hat{U}_\mathrm{pol}^{\dagger} \hat{H}_\mathrm{d \cdot E}^{'}\hat{U}_\mathrm{pol}  \hat{U}_\mathrm{pol}^{' \dagger}) U_{\phi} \nonumber \\
 &= \hat{ T}_{\bf Q} + \mathcal{\hat{H}}_{el}   +  \sum_{{\boldsymbol k}, {\bf\xi}} \Big(  \hat{a}_{{\boldsymbol k}, {\bf\xi}}^{\dagger}  +  \sum_u e (\hat{\bf Q}_u - {\bf R}_u \sum_{\alpha} \hat{c}^{\dagger}_{u, \alpha} \hat{c}_{u, \alpha} )\mathcal{A}^{*}_{{\boldsymbol k},{\boldsymbol \xi}} ({\bf R}_u)  - \sum_{u,\alpha,\alpha'} \mathcal{A}^{*}_{{\boldsymbol k},{\boldsymbol \xi}} ( {\bf R}_u){\boldsymbol \mu}_{\alpha,\alpha'} \hat{c}^{\dagger}_{u, \alpha} \hat{c}_{u, \alpha'} \Big)\nonumber \\
&~~~~~~~~~~~~~~~~~~~~\times\Big(  \hat{a}_{{\boldsymbol k}, {\boldsymbol{\xi}}}  +  \sum_u e(\hat{\bf Q}_u - {\bf R}_u \sum_{\alpha} \hat{c}^{\dagger}_{u, \alpha} \hat{c}_{u, \alpha} )\mathcal{A}_{{\boldsymbol k},{\boldsymbol \xi}} ({\bf R}_u) - \sum_{u,\alpha,\alpha'} \mathcal{A}_{{\boldsymbol k},{\boldsymbol \xi}} ( {\bf R}_u){\boldsymbol \mu}_{\alpha,\alpha'} \hat{c}^{\dagger}_{u, \alpha} \hat{c}_{u, \alpha'}\Big)\omega_{\boldsymbol k}~~~.
\end{align}

We obtain the `electronic' part of the d.E Hamiltonian, so called polaritonic d.E Hamiltonian $\mathcal{\hat{H}}_\mathrm{d \cdot E}(\{{\bf R}_{u}\})$ at the nuclear configuration $\{{\bf R}_{u} \}$  as
\begin{align}
 \mathcal{\hat{H}}_\mathrm{d \cdot E}(\{{\bf R}_{u}\}) =  \mathcal{\hat{H}}_{el}    +  &\sum_{{\boldsymbol k}, {\bf\xi}} \Big(  \hat{a}_{{\boldsymbol k}, {\bf\xi}}^{\dagger}  + \sum_u e ({\bf R}_u - {\bf R}_u \sum_{\alpha} \hat{c}^{\dagger}_{u, \alpha} \hat{c}_{u, \alpha} )\mathcal{A}^{*}_{{\boldsymbol k},{\boldsymbol \xi}} ({\bf R}_u)  -  \sum_{u,\alpha,\alpha'} \mathcal{A}^{*}_{{\boldsymbol k},{\boldsymbol \xi}} ( {\bf R}_u){\boldsymbol \mu}_{\alpha,\alpha'} \hat{c}^{\dagger}_{u, \alpha} \hat{c}_{u, \alpha'} \Big)\nonumber \\
&~\times\Big(  \hat{a}_{{\boldsymbol k}, {\boldsymbol{\xi}}}  + \sum_u e({\bf R}_u - {\bf R}_u \sum_{\alpha} \hat{c}^{\dagger}_{u, \alpha} \hat{c}_{u, \alpha} )\mathcal{A}_{{\boldsymbol k},{\boldsymbol \xi}} ({\bf R}_u) - \sum_{u,\alpha,\alpha'} \mathcal{A}_{{\boldsymbol k},{\boldsymbol \xi}} ( {\bf R}_u){\boldsymbol \mu}_{\alpha,\alpha'} \hat{c}^{\dagger}_{u, \alpha} \hat{c}_{u, \alpha'}\Big)\omega_{\boldsymbol k}~~~.
\end{align}


\section{\normalsize  Exciton-Polariton Hamiltonian} We construct a simple (Frenkel) exciton-polariton Hamiltonian using the d.E Hamiltonian provided in the previous section. We adopt a simple two-band model with dispersionless valence and conduction bands, separated in energy by $\omega_0$. We use a matter basis containing a ground state $|G\rangle = \prod_{u} \hat{c}_{u,g}^{\dagger} |0\rangle$ where all electrons fill up the valence band and localized excitonic states $|E_u\rangle =  \hat{d}_u^\dagger  |G\rangle = \hat{c}_{u,e}^{\dagger}\hat{c}_{u,g}|G\rangle$ where one electron is placed in a localized excited state from a localized ground state at ${\bf R}_u$\cite{BassaniINCD1986,Combescot2008PRB}. These localized excitons interact via nearest-neighbor interactions as is done in the standard Frankel exciton model. We also assume that within this basis, the permanent dipoles $\mu_{\alpha,\alpha} = 0$ and only retain the transition dipoles $\mu_{\alpha,\alpha'}$ with $\alpha \ne \alpha'$. Further, using a transformation   $U_{\phi} = \prod_{{\boldsymbol k},{\boldsymbol \xi}} e^{i\frac{\pi}{2}\hat{a}_{{\boldsymbol k},{\boldsymbol \xi}}^{\dagger}\hat{a}_{{\boldsymbol k},{\boldsymbol \xi}}}$ such that $U_{\phi}^{\dagger} \hat{a}_{{\boldsymbol k},{\boldsymbol \xi}} U_{\phi} = -i \hat{a}_{{\boldsymbol k},{\boldsymbol \xi}}$ (which makes the light-matter couplings real) we obtain the following exciton-polariton dipole gauge Hamiltonian (dropping the nuclear kinetic energy operator) at the nuclear configuration $\{\hat{\bf Q}_{u}\} = \{{\bf R}_{u}\}$,
\begin{align}\label{eqn:HdE0}
\mathcal{\hat{H}}_\mathrm{d \cdot E}     &=   \omega_0 \sum_u \hat{d}_{u}^{\dagger}\hat{d}_{u}  + \sum_{\langle u , v\rangle} \tau_{uv} (\hat{d}_{u}^{\dagger}\hat{d}_{v} + \hat{d}_{u}\hat{d}_{v}^{\dagger})    + \sum_{{\boldsymbol k}, u, {\boldsymbol \xi}}   (\hat{\boldsymbol \mu}_u\cdot \hat{\boldsymbol \lambda}_{{\boldsymbol k},{\boldsymbol \xi}})  \Big(e^{- i {{ \bf k_\parallel}} \cdot {\bf R}_{u}} \hat{a}_{{\boldsymbol k},{\boldsymbol \xi}}^\dagger +  e^{ i { \bf k_\parallel} \cdot {\bf R}_{u}} \hat{a}_{{\boldsymbol k},{\boldsymbol \xi}} \Big) \sin({ k_y} (\hat{\bf y}\cdot {\bf R}_{u})) \\
&+  \sum_{{\boldsymbol k},{\boldsymbol \xi}} \hat{a}_{{\boldsymbol k},{\boldsymbol \xi}}^{\dagger}\hat{a}_{{\boldsymbol k},{\boldsymbol \xi}}\omega_c({\boldsymbol k}) + \sum_{u,v} \sum_{{\boldsymbol k},{\boldsymbol \xi}}  {\frac{ e^{i { \bf k_\parallel} \cdot ({\bf R}_{v}- {\bf R}_{u})}}{ \omega_c({\boldsymbol k}) }  }     (\hat{\boldsymbol \mu}_u\cdot {\boldsymbol \lambda}_{{\boldsymbol k},{\boldsymbol \xi}})  (\hat{\boldsymbol \mu}_v\cdot {\boldsymbol \lambda}_{{\boldsymbol k},{\boldsymbol \xi}})\sin({ k_y} (\hat{\bf y}\cdot {\bf R}_{u}))\sin({ k_y} (\hat{\bf y}\cdot {\bf R}_{v}))\nonumber~~,
\end{align}

where the last two lines describe the dipole self-energy term (DSE), and $\langle u, v\rangle$ implies the summation over the nearest neighbors within the same layer. 
Here we have assumed that the transition dipole is aligned in the $\hat{\bf z}$ direction such that $(\hat{\boldsymbol \mu}_{u} \cdot  \hat{\boldsymbol e}_{{\boldsymbol k}, {s}}) =  \mu_0(\hat{d}_{u,z}^{\dagger} + \hat{d}_{u,z})$  and $(\hat{\boldsymbol \mu}_{u} \cdot  \hat{\boldsymbol e}_{{\boldsymbol k}, {p}})  = 0$.  With this simplification, we can ignore all the $p$ polarized cavity modes and write a simple dipole-gauge Hamiltonian.  We  drop the labels $s$ or $z$, such that $\hat{a}_{{\boldsymbol k}, s} \rightarrow \hat{a}_{{\boldsymbol k}}$   to reduce redundancy and write the following tight-binding (polarization-less) dipole gauge Hamiltonian

 \begin{align} \label{PF}
\mathcal{\hat{H}}_\mathrm{d \cdot E}  &= \sum_{{\boldsymbol k}} \hat{a}_{{\boldsymbol k}}^{\dagger}\hat{a}_{{\boldsymbol k}}\omega_c({\boldsymbol k}) + \omega_0 \sum_u \hat{d}_{u}^{\dagger}\hat{d}_{u}  + \sum_{\langle u , v\rangle} \tau_{uv} (\hat{d}_{u}^{\dagger}\hat{d}_{v} + \hat{d}_{u}\hat{d}_{v}^{\dagger})
 +\sum_{{\boldsymbol k}, u}  \mu_0 \lambda_{\boldsymbol k} (\hat{d}_{u}^{\dagger} + \hat{d}_{u})\big(e^{- i {  \boldsymbol k_\parallel} \cdot {\bf R}_{u}} \hat{a}_{{\boldsymbol k}}^\dagger +  e^{ i {\boldsymbol k_\parallel} \cdot {\bf R}_{u}} \hat{a}_{{\boldsymbol k}} \big) \sin({ k_y} (\hat{\bf y}\cdot {\bf R}_{u}))\nonumber\\
&~+  \sum_{u,v, {\boldsymbol k}}  {\frac{  \mu_0^2 \lambda_{\boldsymbol k}^2 }{{\omega_{\boldsymbol k}}}} (\hat{d}_{u}^{\dagger} + \hat{d}_{u}) (\hat{d}_{v}^{\dagger} + \hat{d}_{v})  e^{i {\bf k}_\parallel \cdot ({\bf R}_{v}- {\bf R}_{u})}  \sin({ k_y} (\hat{\bf y}\cdot {\bf R}_{u}))\sin({ k_y} (\hat{\bf y}\cdot {\bf R}_{v})).  
\end{align}
Note that the above Hamiltonian can be generalized to {\it ab initio} systems where there are multiple energetically relevant electronic transitions with transition dipole moments aligned along arbitrary directions such that both $s$ and $p$ polarization need to be taken into account.


We can further apply additional approximations to simplify the above Hamiltonian. First, we perform the rotating wave approximation (dropping   $\hat{a}_{\boldsymbol k}^{\dagger}\hat{d}_{u}^{\dagger}$ and $\hat{a}_{\boldsymbol k} \hat{d}_{u} $) and drop  the dipole self-energy term to obtain the following generalized Tavis-Cummings Hamiltonian~\cite{Keeling2020ARPC,Mandal2022CR} (GTC)  as
\begin{align}
&\mathcal{\hat{H}}_\mathrm{GTC} = \sum_{{\boldsymbol k}} \hat{a}_{\boldsymbol k}^{\dagger}\hat{a}_{\boldsymbol k}\omega_c({\boldsymbol k}) + \omega_0 \sum_u \hat{d}_{u}^{\dagger}\hat{c}_{u}  
+ \sum_{\langle u , v\rangle} \tau_{uv} (\hat{d}_{u}^{\dagger}\hat{d}_{v} + \hat{d}_{u}\hat{d}_{v}^{\dagger})  + \sum_{{\boldsymbol k}, u}  \mu_0 \lambda_{\boldsymbol k}  \big(e^{- i { \boldsymbol k_\parallel} \cdot {\bf R}_{u}} \hat{a}_{{\boldsymbol k}}^\dagger \hat{d}_{u} +  e^{ i {\boldsymbol  k_\parallel} \cdot {\bf R}_{u}} \hat{a}_{{\boldsymbol k}}\hat{d}_{u}^{\dagger} \big)\sin({ k_y}  (\hat{\bf y}\cdot {\bf R}_{u})). 
 \label{TC}
\end{align}
We investigate the validity of the approximations used to obtain the generalized Tavis-Cummings Hamiltonian below in Sec.~\ref{Sec:Mode-Trunc}. Importantly, we show that the $\mathcal{\hat{H}}_\mathrm{GTC}$ remains valid only when considering a few energetically relevant cavity modes.
 \begin{figure*}[!t]
\centering
\includegraphics[width=0.7\linewidth]{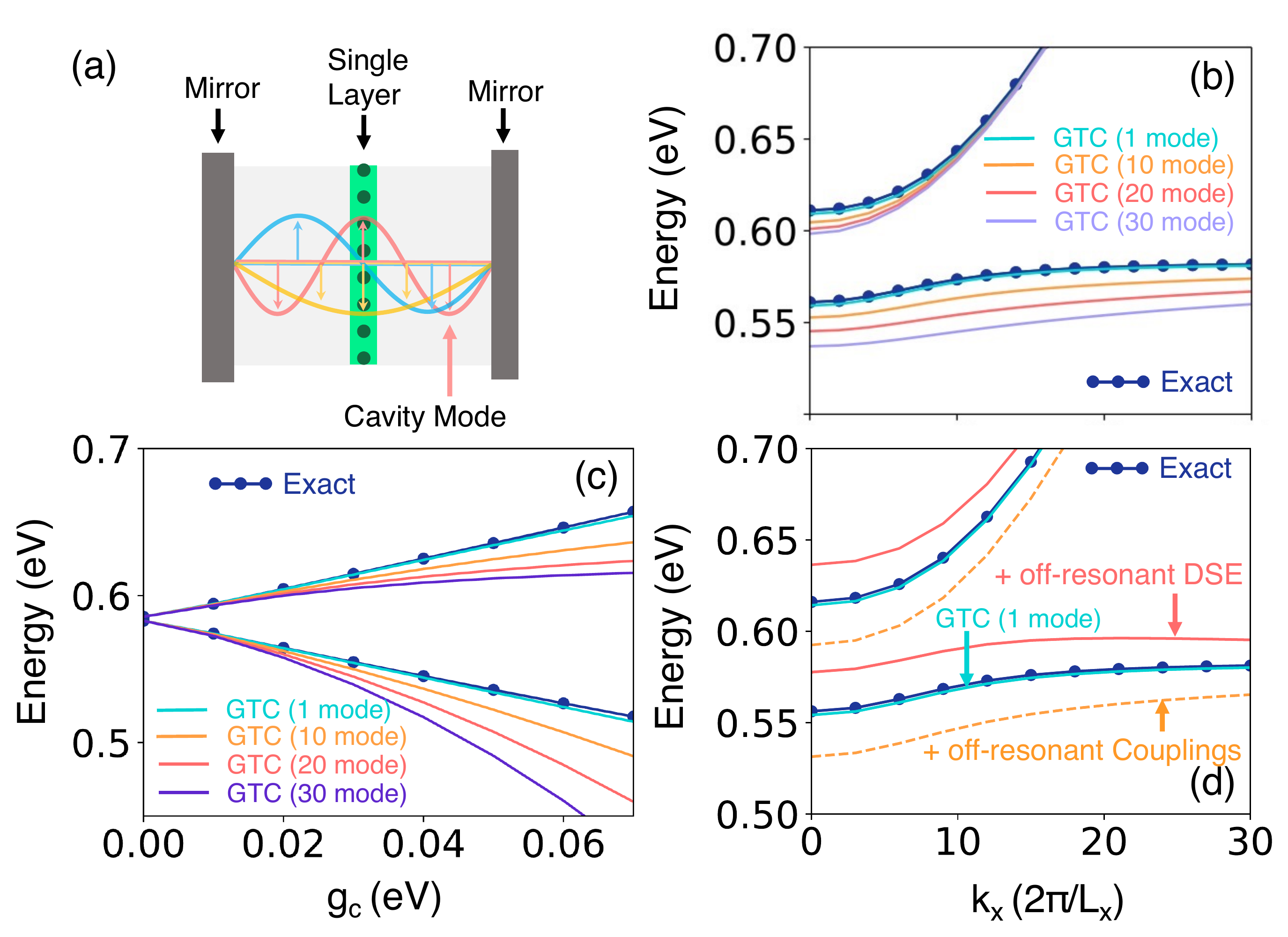}
\caption{\footnotesize (a) Schematic illustration of a single layer material coupled to cavity radiation.  (b) Exact band structure (dark blue) computed with the full dipole gauge Hamiltonian $\hat{\mathcal{H}}_\mathrm{d.E}$ in Eqn.~\ref{PF} compared with the approximate band  structure computed using $\hat{\mathcal{H}}_\mathrm{GTC}$ considering different number of cavity mode branches. (c) Same as (b) but plotting the polariton bands at $k_x = 0$ as a function of light-matter coupling. (d) Polariton bands computed from  $\hat{H}_\mathrm{d.E}$  (exact) compared with the case when only including the dipole-self energy terms of 4 off-resonant cavity modes (total of five cavity modes) and when only including the coupling term between these off-resonant modes and the matter. Here we used $L_y = 20000$ a.u.  and the refractive index $\eta = 1$.}
\label{fig0}
\end{figure*}

\subsection{\normalsize d.E Hamiltonian at $k_z = 0$}
We place the localized excitations in a grid with positions ${\bf R}_{n,m,l} = X_{n} \hat{\bf x} + Y_{m} \hat{\bf y} + Z_l \hat{\bf z}$. To simplify our considerations, we will only consider nearest-neighbor coupling, such that $\hat{d}_{n,m,l}^{\dagger}$ couples to $\hat{d}_{n\pm 1,m,l}$,  $\hat{d}_{n,m\pm 1,l}$ and $\hat{d}_{n,m,l\pm 1}$ through $\tau_x$,  $\tau_y$ and $\tau_z$, respectively. Since here we focus on multi-layer materials where the inter-layer coupling is weak or negligible, we set $\tau_y = 0$.  

Next, we consider periodic boundary conditions along $\hat{x}$ and $\hat{z}$ directions, such that   ${\bf R}_{N+1,m,l} \equiv {\bf R}_{1,m,l}$ and  ${\bf R}_{n,m,N+1} \equiv {\bf R}_{n,m,1}$. Due to the periodic boundaries in the $\hat{x}$ and $\hat{z}$ directions, 
 the cavity modes ${\boldsymbol{k}} = (k_x, k_y, k_z)$ are quantized such that $k_x = \frac{2\pi}{L_x}n_x$  with  $n_x = 0, \pm 1, \pm  2, ..., \pm n_{x_\text{max}}$ and $k_z = \frac{2\pi}{L_z}n_z$ with $n_z = 0, \pm 1, \pm  2, ..., \pm n_{z_\text{max}}$.~\cite{Tichauer2021JCP} Here $L_x$ and $L_z$ are length of the periodic super-cell along $x$ and $z$ direction respectively. The unit-cell along the $\hat{\bf x}$ and $\hat{\bf z}$  direction are separated by $\delta l_x = \delta l_z$ such that $X_{n} = n\cdot \delta l_x$ with $n = 1, 2, ..., N_x$ with a similar expression for $Z_{l}$. 

In most experiments, including ours, the polariton dispersion is plotted along $k_x$ while $k_z$ is set to 0. To obtain this dispersion, we define $\hat{d}_{k_x, m, k_z} = \frac{1}{\sqrt{N_x}} \sum_{n} \hat{d}_{n, m, k_z}e^{-i k_z  X_{n}}    = \frac{1}{\sqrt{N_x N_z}}\sum_{n, l}   e^{-i (k_x  X_{n} + k_z Z_{l})} \hat{d}_{n,m,l}$ and use it to transform the d.E Hamiltonian such that  it is block diagonalized into $|k_z|$ subspaces. Below we write the $k_z = 0$ block (which is decoupled from $k_z \ne 0$) of the d.E Hamiltonian as
\begin{align} \label{exact}
\mathcal{\hat{H}}_\mathrm{d \cdot E}(k_z=0) &= \sum_{{ k_x, k_y}} \hat{a}_{\boldsymbol k}^{\dagger}\hat{a}_{\boldsymbol k}\omega_c({\boldsymbol k}) +  \sum_{k_x,m} \epsilon(k_x) \cdot \hat{d}^{\dagger}_{k_x,m}\hat{d}_{k_x,m}  +  \sum_{{k_x, k_y}, m} g_{\boldsymbol k}   \big(  d_{k_x,m} \hat{a}_{{\boldsymbol k}}^\dagger +     d_{k_x,m}^\dagger\hat{a}_{{\boldsymbol k}} + d_{-k_x,m}^\dagger \hat{a}_{{\boldsymbol k}}^\dagger +    d_{ -k_x, m}\hat{a}_{{\boldsymbol k}}\big) \sin({ k_y} Y_m) \nonumber\\
&~+  \sum_{m,m'}\sum_{k_x, k_y}  {\frac{ g_{\boldsymbol k}^2}{{\omega_{\boldsymbol k}}}} (d_{k_x,m}^{\dagger}  + d_{-k_x, m})( d_{ -k_x, m'}^{\dagger}+ d_{k_x, m'}  )        \sin({ k_y}  Y_{m'})\sin({ k_y} Y_m),
\end{align}
where   we have simplified the indexes such that  $\hat{d}_{k_x,m, k_z = 0} \equiv \hat{d}_{k_x,m}$, we have defined $g_{\boldsymbol k} = \sqrt{N_x N_z}\mu_0 \lambda_{\boldsymbol k} $ and
 $\epsilon(k_x) = \omega_{0} - 2\tau_z- 2\tau_x\cos(k_x  \delta l_x)$. The above Hamiltonian is block diagonalized into $|k_x|$ subspaces each containing $\{\hat{a}_{k_x,k_y}, \hat{a}_{-k_x,k_y}, \hat{d}_{k_x,m}, \hat{d}_{-k_x,m}\}$ and their Hermitian conjugates. A simplified form of the Hamiltonian, the generalized Tavis-Cummings Hamiltonian, is obtained by performing the RWA and dropping the dipole self-energy (DSE) term
\begin{align}\label{TC-k.x}\nonumber
\mathcal{\hat{H}}_\mathrm{GTC}(k_z=0)
&= \sum_{k_x} 
\Bigg( 
\sum_{k_y}\hat{a}_{\boldsymbol k}^{\dagger}\hat{a}_{\boldsymbol k}\omega_c({\boldsymbol k}) + \sum_{m}\epsilon(k_x)\hat{d}^{\dagger}_{k_x,m}\hat{d}_{k_x,m}  + \sum_{{k_y},m}g_{\boldsymbol k}   \big(  \hat{a}_{{\boldsymbol k}}^\dagger \hat{d}_{k_x,m} +    \hat{a}_{{\boldsymbol k}} \hat{d}_{ k_x,m}^{\dagger} \big)\sin({ k_y}  Y_{m})\Bigg) \\
&\equiv\sum_{k_x} \hat{h}_\mathrm{GTC}(k_x) ~~,
\end{align}
which is block-diagonal in each $k_x$ subspace containing only $\{\hat{a}_{k_x,k_y},   \hat{d}_{k_x,m}\}$
and their Hermitian conjugates with ${\hat{h}}_\mathrm{GTC}(k_x)$ as the Hamiltonian in the $k_x$ block.  In this work, we use  $\mathcal{\hat{H}}_\mathrm{GTC}(k_z = 0)$ due to its simple block-diagonalized structure, as opposed to $\mathcal{\hat{H}}_\mathrm{d.E}(k_z = 0)$, and develop simple matrix models to obtain the polariton dispersion for multi-layered material inside the cavity.  

\section{\normalsize Mode truncation and dipole self-energy}\label{Sec:Mode-Trunc}
Here we discuss how to deal with energetically off-resonant cavity modes in the GTC Hamiltonian  $\mathcal{\hat{H}}_\mathrm{GTC}$ and comment on the role of missing DSE terms.  
To assess the accuracy of  $\mathcal{\hat{H}}_\mathrm{GTC}(k_z = 0)$, we consider a single layer material placed inside an optical cavity as depicted in Fig.~\ref{fig0}a.  The exact polariton dispersion is computed by solving $\mathcal{\hat{H}}_\mathrm{d.E}(k_z = 0)$ given in Eqn.~\ref{exact}, and is shown as dark blue solid lines with filled circles in Fig.~\ref{fig0}b-d. 
These exact results are computed by converging with respect to the number of cavity mode along $k_y$ and the number of Fock states for each of these modes. In Fig.~\ref{fig0} we use $\tau_x = 150$ cm$^{-1}$ and $\omega_0 - 2\tau_z = 0.62$ eV. 


When using the $\mathcal{\hat{H}}_\mathrm{d.E}(k_z = 0)$,  our result shows that the energetically off-resonant cavity modes do not contribute to the low-energy polaritonic subspace, as one expects.  
However, counterintuitively, the contributions to the DSE term these off-resonant modes and the associated light-matter coupling term are non-negligible. DSE makes a positive contribution, while the coupling term makes a negative contribution. Thus, when  performing mode truncation by removing cavity modes with $k_y> k_{y_\mathrm{max}}$  that are energetically off-resonant in $\mathcal{\hat{H}}_\mathrm{d.E}(k_z = 0)$, both the corresponding coupling terms as well as the associated  DSE terms must be dropped. This means we must restrict the sum $\sum_{k_y} \rightarrow \sum_{k_y \le k_{y_\text{max}}}$ for both the coupling terms as well as the DSE terms despite the DSE being purely a matter operator (which ensures gauge invariance~\cite{Taylor2022OL,Taylor2020PRL}). 

Therefore, when considering $\mathcal{\hat{H}}_\mathrm{GTC}(k_z = 0)$ which ignores the dipole self-energy but retains the light-matter coupling term partially (since we employ the RWA), this balance is broken. As a result, a very undesirable feature of $\mathcal{\hat{H}}_\mathrm{GTC}(k_z = 0)$ (and thus the $\mathcal{\hat{H}}_\mathrm{GTC}$) is that the polariton eigenspectrum (or polariton dispersion) computed with it does not converge with respect to the number of cavity mode $k_y$. This is shown in Fig.~\ref{fig0}(b)-(d). 

In Fig.~\ref{fig0}(b), we present the polariton band structure computed exactly using direct diagonalization of $\mathcal{\hat{H}}_\mathrm{d.E}(k_z = 0)$ compared with the approximate $\mathcal{\hat{H}}_\mathrm{GTC}(k_z = 0)$ using different numbers of $k_y$ cavity modes. Here we choose a light-matter coupling   $g_c = 25$ meV where $g_\mathrm{k} = \sqrt{\frac{\omega_c({k_x,k_y})} {\omega_c(0,\pi/L_y)}}g_c$ (note that $k_z = 0$ and $\omega_c({k_x,k_y}) \equiv \omega_c({k_x,k_y, 0}) = c \sqrt{k_x^2 + k_y^2}$). When considering only one mode $\mathcal{\hat{H}}_\mathrm{GTC}(k_z = 0)$ provides visually indistinguishable  results compared to the exact benchmark. However, as number of $k_y$ cavity modes are increased  in  $\mathcal{\hat{H}}_\mathrm{GTC}(k_z = 0)$ the cavity dispersion starts to diverge. The same is observed in  Fig.~\ref{fig0}(c), where we compute the polariton bands at $k_x = 0$ as a function of the coupling $g_c$. This result corroborates previous work in the single-molecule cavity setup~\cite{Taylor2022OL} and in circuit QED~\cite{Malekakhlagh2017PRL} under the long-length approximation. Note that here, importantly, in contrast to these other works, we only make the long wavelength approximation on the atomic scale, not on the scale of the entire materials.

To illustrate the cancellation of the DSE and coupling terms for the off-resonant cavity modes, in  Fig.~\ref{fig0}(d), we considered eight cavity mode branches $k_y \in \{\frac{\pi} {L_y},\frac{2\pi} {L_y}, ..., \frac{8\pi} {L_y}\}$ where $k_y = \frac{\pi} {L_y}$ mode is resonant with matter excitations at $k_x = 0$ (the rest of the modes can be regarded as off-resonant) with $g_c = 30$ meV. We compare the full dipole gauge Hamiltonian (i.e. `exact') with an approximation where we remove all the dipole self-energy (DSE) terms associated with off-resonant cavity modes $k_y > \frac{\pi} {L_y}$ while retaining the coupling terms between matter and off-resonant cavity modes (cyan dashed lines). Finally, we compute the dispersion when including the DSE terms associated with the off-resonant cavity modes ($k_y > \frac{\pi} {L_y}$), while removing the coupling terms between matter and off-resonant cavity modes (red solid lines). As explained before, it can be seen that the DSE terms (coming from the off-resonant modes) contribute a positive amount while the coupling terms contribute a negative amount, and the magnitude of these two contributions is the same (up to the second order in light-matter coupling as shown in Ref.~\cite{Taylor2022OL}) but with opposite signs. When including both, their overall contribution (originating from the off-resonant modes) vanishes, and thus a single-mode GTC generates accurate results compared to exact. Therefore, when using the GTC model, we must restrict our consideration to a few energetically relevant cavity modes and, the (conveniently) drop all energetically off-resonant modes from the Hamiltonian. 
}

\section{\normalsize  Various cavity setups and corresponding matrix models}

\subsection{\normalsize Single Layer}
{\footnotesize 
First, we consider the simplest scenario, a single layer, such that the sum over $m$ and the index itself can be removed (such that $\hat{d}_{k_x, m} \rightarrow   \hat{d}_{k_x}$). The ${\hat{h}}_\mathrm{GTC}(k_x)$ Hamiltonian for a single layer is then written as,
\begin{align}
{\hat{h}}_\mathrm{GTC}(k_x) &= \sum_{{  k_y}} \hat{a}_{\boldsymbol k}^{\dagger}\hat{a}_{\boldsymbol k}\omega_c({\boldsymbol k})+  \epsilon(k_x)\hat{d}^{\dagger}_{k_x}\hat{d}_{k_x}     +   \sum_{k_y}{g_{\boldsymbol k}}    \big(  \hat{a}_{{\boldsymbol k}}^\dagger \hat{d}_{k_x} +    \hat{a}_{{\boldsymbol k}} \hat{d}_{ k_x}^{\dagger} \big)\sin({ k_y}  Y_{0}), 
\end{align}
where $\epsilon(k_x) = \omega_{0} - 2\tau_x\cos(k_x  \delta l_x)$, $Y_0$ is the location of the single layer along the $\hat{\bf y}$ direction inside the cavity. For this single layer placed between the two mirrors  $0 < Y_0 < L_y$, the ${\hat{h}}_\mathrm{GTC}(k_x)$ Hamiltonian can be written conveniently in the single excited subspace using the basis spanning $\{\hat{a}_{{k_x, k_y}}^\dagger| G,0\rangle,  \hat{d}_{k_x}^{^\dagger}|G,0\rangle \}$ where $|G,0\rangle$ represents the matter ground state with no photons in the cavity. 
For a given $k_x$, this is the basis of $N$ cavity modes to account for $k_y$ and one matter state for the exciton.
Therefore, ${\hat{h}}_\mathrm{GTC}(k_x)$ matrix  takes the form of the  widely used $(N+1)\times (N+1)$ matrix~\cite{ Dietrich2016SA, Coles2014APL, Michetti2005PRB, Richter2015APL,Georgiou2021JCP,Balasubrahmaniyam2021PRB,Orosz2011APE,Faure2009APL} 
\begin{align}\label{SL-Hij}
{\hat{h}}_\mathrm{N+1} = \begin{bmatrix}
\epsilon(k_x) &  {\Omega}(\frac{\pi}{L_y})  &  {\Omega}(\frac{2\pi}{L_y}) & \hdots \\
{\Omega}(\frac{\pi}{L_y})  & \omega_c({k_x, \frac{\pi}{L_y}}) & 0 & \hdots\\
{\Omega}(\frac{2\pi}{L_y}) & 0 & \omega_c({k_x, \frac{2\pi}{L_y}} )
& \hdots \\
\vdots & \vdots    & \vdots  & \ddots \end{bmatrix} ~,
\end{align}
 where $ {\Omega}({k_y})  =    g_{\boldsymbol k} \sin(k_y \cdot Y_0)$ and $\epsilon(k_x)  = \omega_{0} - 2\tau_x\cos(k_x  \delta l_x)$. Diagonalizing the above matrix provides the polariton dispersion along $k_x$ for a single layer coupled to cavity radiation modes.

}
\subsection{\normalsize Filled cavity with non-interacting layers}

{\footnotesize

Here, we consider a material filling the cavity ($l_y = L_y$). We assume that $m$th layer is centered at $Y_m = (m-\frac{1}{2})\delta L_y$ where $m = 1, 2, ... N_y$ and $k_y = n_y\frac{\pi}{L_y}$. For this setup, the collective matter excitation operators are found using the unitary matrix $[U_\mathrm{O}]_{m,n_y} = \sqrt{\frac{2}{N_y}}\sin (\frac{\pi}{N_y} n_y\cdot (m-\frac{1}{2}) ) $ for $n_y \ne N_y$
and  $[U_\mathrm{O}]_{m,N_y} = \sqrt{\frac{1}{N_y}}\sin (\frac{\pi}{N_y} N_y\cdot (m-\frac{1}{2}) )$ which is a the discrete sine transform matrix of type II. With the collective matter excitation operators $\hat{d}_{k_x,k_y} = \sum_{m} [U_\mathrm{O}]_{m,n_y} \hat{d}_{k_x,m}$ we obtain  ${\hat{h}}_\mathrm{GTC}(k_x)$. Using the single excited subspace spanning $\{\hat{a}_{{k_x, k_y}}^\dagger| G,0\rangle,  \hat{d}_{{k_x, k_y}}^{^\dagger}|G,0\rangle \}$ where $|G,0\rangle$ represents the matter ground state with no photons in the cavity, we find the following $2N\times 2N$ model 


\begin{align}\label{filled-noint}
{\hat{h}}_{\mathrm{2N}}({k_x}) = 
\left[\!\begin{array}{*{24}{c@{\hspace{4pt}}}}
\epsilon(k_x) &  {\Omega}(\frac{\pi}{L_y})   & 0 & 0 & \hdots \\
{\Omega}(\frac{\pi}{L_y})  & \omega_c({k_x, \frac{\pi}{L_y}}) & 0 &  0 & \hdots\\
0 & 0 & \epsilon(k_x) & {\Omega}(\frac{2\pi}{L_y})
& \hdots \\
0   & 0 & {\Omega}(\frac{2\pi}{L_y}) &   \omega_c({k_x, \frac{2\pi}{L_y}})& \hdots\\
\vdots & \vdots & \vdots   & \vdots  & \ddots \end{array}\!\right]
\end{align}
where $ {\Omega}(k_y)  =   \sqrt{\frac{N_y}{2}} g_{k_x, k_y} \propto  \sqrt{N_x N_y N_z}$ (note that $g_{\bf k} \propto \sqrt{N_x N_z}$) also  ${\Omega}_y  \propto  \sqrt{\omega_c (k_x, k_y)}$ and $\epsilon(k_x)  = \omega_{0} - 2\tau_x\cos(k_x  \delta l_x)$. Diagonalizing the above matrix provides the polariton dispersion along $k_x$ for a multi-layer (non-interacting) material filling up the entire cavity. The same results are obtained when considering interacting layers $\tau_y \ne 0$ (shown in the supporting information). This is because the same transformation also diagonalizes the inter-layer coupling term.

\subsection{\normalsize Filled cavity with interacting layers}
Here we consider a filled cavity with interacting layers. For $\tau_y \ne 0$,  we focus on the light-matter coupling term (within a $k_x$ subspace),
\begin{align}
&\sum_{k_y,m}  {g_{\bf k}}  \mu_0(\hat{c}^{\dagger}_{k_x,m}\hat{a}_{\bf k}  + \hat{c}_{k_x,m}\hat{a}_{\bf k}^{\dagger} )  \sin\Big(y\frac{\pi}{N_y} \cdot \big(m-\frac{1}{2}\big)\Big) \nonumber\\
&\approx  \sum_{k_y \in \mathcal{K}_{c}}  {g_{\bf k}}  \mu_0   \sum_{m}(\hat{c}^{\dagger}_{k_x,m}\hat{a}_{\bf k}  + \hat{c}_{k_x,m}\hat{a}_{\bf k}^{\dagger} )    \Bigg[ \sin\Big(\frac{y\pi m}{N_y}     \Big)\cos( \frac{y\pi}{2N_y}) - \cos\Big(\frac{y\pi m}{N_y}    \Big)\sin\Big( \frac{y\pi}{2N_y}\Big)\Bigg] ~,\nonumber
\end{align}

where in the last-line we confined the sum  to the cavity modes that are energetically relevant, $\mathcal{K}_c = \{\frac{\pi n_\mathrm{min}}{L_y}, \frac{\pi (n_\mathrm{min} + 1)}{L_y}, ...,  \frac{\pi n_\mathrm{max}}{L_y} \}$. For $y\frac{\pi }{2N_y} = \frac{1}{2} k_y \cdot \delta l_y \rightarrow 0$ we have  $\sin\big( y\frac{\pi}{2N_y}\big) \rightarrow 0$ and $\cos\big( y\frac{\pi}{2N_y}\big) \rightarrow 1$, this is equivalently assuming that the wavelength of the relevant cavity modes are much longer than the size of the unit cells (thickness each matter layer). Additionally we assume that $N_y \gg 1$ such that $N_y \approx N_y + 1$ and we write
 \begin{align}
&\sum_{k_y,m}  {g_{\bf k}}  \mu_0  (\hat{c}^{\dagger}_{k_x,m}\hat{a}_{\bf k}  + \hat{c}_{k_x,m}\hat{a}_{\bf k}^{\dagger} )  \sin(k_y \cdot Y_m) \approx  \sum_{k_y \in \mathcal{K}_{c}}  \mu_0{g_{\bf k}}      \sum_{m}(\hat{c}^{\dagger}_{k_x,m}\hat{a}_{\bf k}  + \hat{c}_{k_x,m}\hat{a}_{\bf k}^{\dagger} )   \sin\Big(\frac{\pi y\cdot  m}{N_y+1}  \Big).
\end{align}
We use the discrete sine transform matrix of type I to construct the collective matter operator. Thus we use $[U_\mathrm{O}]_{m,y} = \sqrt{\frac{2}{N_y + 1}}\sin(\frac{y\pi m}{N_y+1})$ such that $\hat{c}_{k_x, k_y} = \sum_{m} [U_\mathrm{O}]_{m,y} \hat{c}_{k_x, m}$. Note that the inter-layer coupling term  is simultaneously diagonalized using these collective matter operators. Thus for cavity filled with multi-layered (with interacting layers) material has the following form of  $\mathcal{\hat{H}}_{k_x} $,
\begin{align} 
 \mathcal{\hat{H}}^{k_x}_\mathrm{GTC}  &= \sum_{ k_y  } \hat{a}_{\boldsymbol k}^{\dagger}\hat{a}_{\boldsymbol k}\omega_c({\boldsymbol k})  +  \Big(\omega_{0} - 2\tau_x\cos(k_x  \delta l_x) - 2\tau_y\cos(k_y  \delta l_y)\Big)\hat{c}^{\dagger}_{k_x, k_y}\hat{c}_{k_x, k_y}     + \mu_0\sqrt{ \frac{N_y + 1} {2}}\sum_{k_y \in \mathcal{K}_c}{g_{\boldsymbol k}}   \big(  \hat{a}_{k_x, k_y}^\dagger c_{ k_x, k_y} +    \hat{a}_{{k_x, k_y}}c_{ k_x, k_y}^{\dagger} \big). \nonumber
  \end{align}
  
 Using the single excited subspace spanning $\{\hat{a}_{{k_x, k_y}}^\dagger| \bar{0}\rangle,  \hat{c}_{{k_x, k_y}}^{^\dagger}|\bar{0}\rangle \}$ gives us the following Hamiltonian matrix (a $2N\times 2N$ model ) that can be diagonalized to obtain the polariton dispersion along $k_x$  
 
\begin{align}\label{filled-int}
 \mathcal{\hat{H}}_{\mathrm{2N}}({k_x}) = \begin{bmatrix}
\epsilon(k_x, \frac{\pi}{L_y}) &  {\Omega}_1    & 0 & 0 & \hdots \\
{\Omega}_1   & \omega_c({k_x, \frac{\pi}{L_y}}) & 0 &  0 & \hdots\\
0 & 0 & \epsilon(k_x, \frac{2\pi}{L_y}) & {\Omega}_2
& \hdots \\
0   & 0 & {\Omega}_2 &   \omega_c({k_x, \frac{2\pi}{L_y}})& \hdots\\
\vdots & \vdots & \vdots   & \vdots  & \ddots \end{bmatrix},
\end{align}
 where $ {\Omega}_y  =  \mu_0\sqrt{\frac{ N_y + 1}{2}} g_{k_x, k_y} \propto \sqrt{\omega_c (k_x, k_y)}$ and $\epsilon(k_x, k_y)  = \omega_{0} - 2\tau_x\cos(k_x  \delta l_x)-  2\tau_y\cos(k_y  \delta l_y)$.

\subsection{\normalsize  Partially filled cavity with thin material placed next to a mirror}
Here we consider a material thickness $l_y \ll L_y$ and material is placed near one of the cavity mirrors. The $m$th layer of the material is placed at $Y_m = (m-\frac{1}{2})\delta L_y$ where $m = 1, 2, ...  N_y$ and for $l_y \ll L_y$ we approximate $\sin( k_y \cdot Y_m) \approx k_y \cdot Y_m$. With this the light-matter coupling term (in $k_x$ subspace)   Eqn.~\ref{TC-k.x} is written as
 \begin{align}
&\sum_{k_y,m}  {g_{\bf k}} (\hat{d}^{\dagger}_{k_x,m}\hat{a}_{\bf k}  + \hat{d}_{k_x,m}\hat{a}_{\bf k}^{\dagger} )  \sin(k_y \cdot Y_m) \approx  \sum_{k_y \in \mathcal{K}_{c}}   {g_{\bf k}}      k_y \sum_{m}(\hat{d}^{\dagger}_{k_x,m}\hat{a}_{\bf k}  + \hat{d}_{k_x,m}\hat{a}_{\bf k}^{\dagger} )   Y_m,
\end{align}
where in the second line we also only consider energetically relevant cavity mode subspace $\mathcal{K}_\mathrm{c}$ as before. We define $[U_\mathrm{O}]_{m,n_\mathrm{min}} = {\frac{1}{\sqrt{\mathcal{N}_B}}}   Y_m$ where $ \mathcal{N}_B = {\frac{N_y(N_y + 1 ) (2N_y +1) }{6}} - \frac{3}{4}N_y$ is a normalization factor. Following the same strategy as before we define a collective $\hat{d}_{k_x, B} = {\frac{1}{\sqrt{\mathcal{N}_B}}}   \sum_{m}Y_m \hat{d}_{k_x,m}$ and the other $N-1$ dark matter excitation operators $ \hat{d}_{k_x, D_y} = \sum_{m}[U_\mathrm{O}]_{m,n_y\ne n_\mathrm{min}} \hat{d}_{k_x,m}$. Since the dark operators as $\{\hat{d}_{k_x, D_y}\}$ are decoupled from the bright matter operator as well as photonic operators, we ignore them and write the following Hamiltonian, 

\begin{align}
{\hat{h}}_\mathrm{GTC}(k_x) &= \sum_{{  k_y  }} \hat{a}_{\boldsymbol k}^{\dagger}\hat{a}_{\boldsymbol k}\omega_c({\boldsymbol k})+  \epsilon(k_x)\hat{d}^{\dagger}_{k_x, B}\hat{d}_{k_x, B}    +  \sum_{k_y \in \mathcal{K}_c} k_y {g_{\boldsymbol k}} \sqrt{\mathcal{N}_B}   \big(  \hat{a}_{k_x, k_y}^\dagger \hat{d}_{ k_x, B} +    \hat{a}_{{k_x, k_y}}\hat{d}_{ k_x, B}^{\dagger} \big)  ~~.  
\end{align}
 Using  the single excited subspace spanning $\{\hat{a}_{{k_x, k_y}}^\dagger| G,0\rangle,  \hat{d}_{{k_x, k_y}}^{^\dagger}|G,0\rangle \}$, we obtain the following $(N+1)\times(N+1)$ model

\begin{align}\label{ML-matrix-bottom}
{\hat{h}}_{\mathrm{N+1}}({k_x})= \begin{bmatrix}
\epsilon(k_x ) &  {\Omega} (\frac{\pi}{L_y})    & {\Omega}(\frac{2\pi}{L_y})    & \hdots \\
{\Omega}(\frac{\pi}{L_y})    & \omega_c({k_x, \frac{\pi}{L_y}}) &  0 & \hdots\\
{\Omega}(\frac{2\pi}{L_y})  & 0 & \omega_c({k_x, \frac{2\pi}{L_y}})  
& \hdots \\
\vdots   & \vdots   & \vdots  & \ddots \end{bmatrix},
\end{align}
 where $ {\Omega}(k_y) =    k_y \sqrt{\mathcal{N}_B} g_{k_x, k_y} \propto k_y\cdot \sqrt{\omega_c (k_x, k_y)}$ and $\epsilon(k_x )  = \omega_{0} - 2\tau_x\cos(k_x  \delta l_x)$.

\subsection{\normalsize Partially filled cavity with thin material placed at the middle}
Similar to the previous section, here we also consider $l_y \ll L_y$  but we place the material in the center of the cavity. Thus   $Y_m = m \cdot \delta l_y + L_y/2 - l_y/2$. With a geometry where $\frac{2\pi n_\mathrm{min}}{k_y} \gg l_y$, we can approximate $\sin(k_y \cdot Y_m) \approx (-1)^{\frac{n_y-1}{2}}$ for odd $n_y = 1, 3, ...$ and $\sin(k_y \cdot Y_m) \approx k_y \cdot (Y_m - L_y/2)$  for even $n_y = 2, 4, ...$ and we introduce a unitary transformation matrix $U_{\mathrm{O}}$ that is of $N_y\times2$ dimensions. We set  $[U_{\mathrm{O}}]_{m,n_\mathrm{min}} = 1/\sqrt{N_y}$ and $[U_{\mathrm{O}}]_{m,n_\mathrm{min} +1 } = (Y_m - L_y/2)/\sqrt{\sum_j (Y_m - L_y/2)^2} = (Y_m - L_y/2)/\sqrt{\mathcal{N}}$ for which we have $\sum_m [U_{\mathrm{O}}]_{m, n_\mathrm{min}} \cdot [U_{\mathrm{O}}]_{m,n_\mathrm{min} + 1} = 0$. With this transformation, we define the symmetric and asymmetric bright  excitonic operators as,
 
 \begin{align}
 \hat{d}^{\dagger}_{k_x, \mathrm{S}} &= \frac{1}{\sqrt{N_y}}  \sum_m  \hat{d}^{\dagger}_{k_x, m},  \\
 \hat{d}^{\dagger}_{k_x, \mathrm{A}} &= \frac{1}{\sqrt{\mathcal{N}}}  \sum_m    \Big(Y_m - \frac{L_y}{2}\Big) \hat{d}^{\dagger}_{k_x, m} ~~.
 \end{align}

Using the $\hat{d}^{\dagger}_{k_x, \mathrm{S}}$ and $\hat{d}^{\dagger}_{k_x, \mathrm{A}}$,  we can write ${\hat{h}}_\mathrm{GTC}(k_x) = {\hat{h}}_{\mathrm{S}}({k_x})  + {\hat{h}}_{\mathrm{A}}({k_x})$ as 
 
\begin{align} 
{\hat{h}}_{\mathrm{S}}(k_x)  &= \sum_{{  k_y \in \mathrm{odd} }} \hat{a}_{\boldsymbol k}^{\dagger}\hat{a}_{\boldsymbol k}\omega_c({\boldsymbol k}) +  \epsilon(k_x)\hat{d}^{\dagger}_{k_x, \mathrm{S}}\hat{d}_{k_x, \mathrm{S}}     + \sqrt{N_y}\sum_{k_y \in \mathrm{odd}}   (-1)^{\frac{n_y-1}{2}} {g_{\boldsymbol k}}    \big(  \hat{a}_{k_x, k_y}^\dagger \hat{d}_{ k_x, \mathrm{S}} +    \hat{a}_{{k_x, k_y}}\hat{d}_{ k_x, \mathrm{S}}^{\dagger} \big)~~,  \\
{\hat{h}}_{\mathrm{A}} ({k_x})  &= \sum_{{  k_y \in \mathrm{even} }} \hat{a}_{\boldsymbol k}^{\dagger}\hat{a}_{\boldsymbol k}\omega_c({\boldsymbol k})  \epsilon(k_x)\hat{d}^{\dagger}_{k_x, \mathrm{A}}\hat{d}_{k_x, \mathrm{A}}  +  \sqrt{\mathcal{N}}\sum_{k_y \in \mathrm{even}} k_y {g_{\boldsymbol k}}   \big(  \hat{a}_{k_x, k_y}^\dagger \hat{d}_{ k_x, \mathrm{A}} +    \hat{a}_{{k_x, k_y}}\hat{d}_{ k_x, \mathrm{A}}^{\dagger} \big)~~.  
\end{align}

The Hamiltonian matrix using a single excited subspace spanning $\{\hat{a}_{{k_x, k_y}}^\dagger| G,0\rangle,  \hat{d}_{{k_x, k_y}}^{^\dagger}|G,0\rangle \}$ is given by a $(N+2)\times (N+2)$ model 
\begin{equation}\label{ML-middle}
    {\hat{h}}_{\mathrm{N+2}}({k_x}) = {\hat{h}}_{\mathrm{S}}({k_x}) \oplus {\hat{h}}_{\mathrm{A}}({k_x})
\end{equation}
 where ${\hat{h}}_{\mathrm{S}}({k_x})$ and ${\hat{h}}_{\mathrm{A}}({k_x})$ are given as
 
\begin{align}\label{ML-middle-sym}
{\hat{h}}_{\mathrm{S}}({k_x}) =
 \begin{bmatrix}
\epsilon(k_x ) &  {\Omega}(\frac{\pi}{L_y})    & {\Omega}(\frac{3\pi}{L_y}) & \hdots \\
{\Omega}(\frac{\pi}{L_y})  & \omega_c({k_x, \frac{\pi}{L_y}}) & 0 &  \hdots  \\
{\Omega}(\frac{3\pi}{L_y})   & 0 & \omega_c({k_x, \frac{3\pi}{L_y}})&  \hdots   \\
\vdots   & \vdots & \vdots &   \ddots  \end{bmatrix} ~~,
\end{align}

\begin{align} 
{\hat{h}}_{\mathrm{A}}({k_x}) =
 \begin{bmatrix}
\epsilon(k_x ) &  {\Omega}(\frac{2\pi}{L_y})  & {\Omega}(\frac{4\pi}{L_y}) & \hdots \\
{\Omega}(\frac{2\pi}{L_y})  & \omega_c({k_x, \frac{2\pi}{L_y}}) & 0 &  \hdots  \\
{\Omega}(\frac{4\pi}{L_y})  & 0 & \omega_c({k_x, \frac{4\pi}{L_y}})&  \hdots   \\
\vdots   & \vdots & \vdots &   \ddots  \end{bmatrix}~~, 
\end{align}

 where $ {\Omega}(k_y)  =     \sqrt{  {N}_y} g_{k_x, k_y} (-1)^{\frac{n_y-1}{2}}  \propto   \sqrt{\omega_c (k_x, k_y)}$ for odd $n_y$, ${\Omega}(k_y)  =  \mu_0  \sqrt{ \mathcal{N}} g_{k_x, k_y}  k_y  \propto   k_y \cdot  \sqrt{\omega_c (k_x, k_y)}$ for even $n_y$ and $\epsilon(k_x )  = \omega_{0} - 2\tau_x\cos(k_x  \delta l_x)$. Note that here the Hamiltonian matrix take the form of a $(N+2)\times (N+2)$ matrix which when only considering two cavity mode branches appear as an $2N\times 2N$ matrix.

}

\section{\normalsize  Experimental Details }

 \subsection{\normalsize CH$_3$(CH$_2$)$_3$NH$_3$)$_2$(CH$_3$NH$_3$)$_2$Pb$_3$I$_{10}$  crystal synthesis}.  {\footnotesize
 
 The synthesis followed an aqueous solution crystallization procedure. Lead (II) iodide (PbI2, 99.999\% trace metals bases) powder (2720 mg, 5.9 mmol) was dissolved in   hydriodic acid solution (HI, 9 mL, 57\% wt.) with 1 mL hypophosphorous acid (H3PO2, 50\% wt.) as the reducing agent. The solution was heated to 105$^o$ C under constant magnetic stirring. Addition of methylammonium iodide (MAI, $\ge$ 99\%, anhydrous 636 mg, 4.0 mmol) and n-butylammonium iodide (BAI, 382 mg, 1.9 mmol) produced black precipitation instantly and redissolved in 5 mins. The solution was subjected to a controlled cooling rate of 0.5 $^\mathrm{o}\mathrm{C}$/h to room temperature in an oil bath. The crystals were collected by vacuum filtering and washed twice with toluene. All chemicals were purchased from Sigma-Aldrich.

 \subsection{\normalsize  Perovskites cavities fabrication.}  The three cavity samples in the main text all consist of exfoliated single crystals of CH$_3$(CH$_2$)$_3$NH$_3$)$_2$(CH$_3$NH$_3$)$_2$Pb$_3$I$_{10}$ placed between a pair of metallic mirrors. The bottom mirror, a 30 nm gold film, was deposited on cover glass (Richard-Allan Scientific, 24×50 \#1.5) by electron-beam evaporation (Angstrom EvoVac deposition system) with a deposition rate of 0.05 nm/s. This thin glass/Au mirror allows partial light transmission for spectroscopic interrogation. The top mirror was a 200 nm silver deposited by the same method. 
For the partially filled cavity of Figure 4b, thin perovskite flakes were mechanically exfoliated with PVC tape (Nitto SPV224 PVC Vinyl Surface Protection Specialty Tape) and transferred onto the bottom Au mirror. A 1040 nm thick layer of PMMA [Poly(methyl methacrylate), 950 MW, Kayaku Advanced Materials] was then deposited on the perovskite through spin-coating at 1000 rpm. The structure is then baked at 170°C for 5 mins to homogenize the interface, before depositing the top silver mirror. 

For the partially filled cavity of Figure 4c, a 515 nm PMMA spacer was spin-coated at 3000 rpm on the bottom Au mirror. The structure is baked at 170°C for 5 mins before exfoliating perovskite single crystals on top of the PMMA. A second PMMA spacer was subsequently spin-coated (3000 rpm, baked at 130°C for 2 mins) on the perovskite flakes, followed by deposition of the top silver mirror.

For the filled cavity sample of Figure 4a, the perovskite flakes were exfoliated onto the bottom Au mirror with PVC tape. The top silver mirror was evaporated on a Si substrate, then template-stripped and deposited on the Au/perovskite structure using thermal release tape.

 }

\bibliographystyle{naturemag}
\bibliography{bib.bib}